\documentclass[a4paper]{jpconf}
\usepackage{graphicx}
\begin{document}
\title{On representations of Higher Spin symmetry  algebras for
mixed-symmetry HS fields on AdS-spaces. Lagrangian formulation}

\author{\v{C}. Burd\'{\i}k${}^{a}$ and A. Reshetnyak${}^{b}$ }

\address{${}^a$Department of Mathematics, Czech Technical University, Prague  12000,  Czech
Republic,\\
${}^{b}$Institute of  Strength Physics and Materials Science,
Tomsk  634021, Russia}

\ead{${}^a$burdik@kmlinux.fjfi.cvut.cz,
${}^b$reshet@ispms.tsc.ru}

\begin{abstract}We derive non-linear commutator HS
symmetry algebra,  which encode unitary irreducible
representations of AdS group subject to Young tableaux
$Y(s_1,\ldots ,s_k)$ with $k\geq 2$ rows on $d$-dimensional
anti-de-Sitter space.  Auxiliary  representations for specially
deformed non-linear HS symmetry algebra in terms of generalized
Verma module in order to additively convert a subsystem of
second-class constraints in the HS symmetry algebra into one with
first-class constraints are found explicitly for the case of HS
fields for $k=2$ Young tableaux. The oscillator realization over
Heisenberg  algebra for obtained Verma module is constructed. The
results generalize the method of auxiliary representations
construction for symplectic $sp(2k)$ algebra used for
mixed-symmetry HS fields on a flat spaces and can be extended on a
case of arbitrary HS fields in AdS-space. Gauge-invariant
unconstrained reducible  Lagrangian formulation for free  bosonic
HS fields with generalized spin $(s_1,s_2)$ is
derived.\end{abstract}

\section{Introduction}
Increased interest to higher-spin field theory is mainly
conditioned by expected output of LHC on the planned capacity. It
suspects not only the proof of supersymmetry display, the answer
on the question on existence of Higgs boson, and possibly a new
insight on origin of Dark Matter (\cite{LHC}, \cite{LHC1}) but
permits one to reconsider the problems of an unique description of
variety of elementary particles and all known interactions. If the
above hopes are true the development of higher-spin (HS) field
theory in view of its close relation to superstring theory on
constant curvature spaces, which operates with an infinite set of
massive and massless bosonic and fermionic HS fields subject to
multi-row Young tableaux (YT) $Y(s_1,...,s_k)$, $k \geq 1$ (see
for a review, \cite{reviews}--\cite{Tsulaia}) seems by actual one.
Corresponding description of such theories, having as the final
aim the Lagrangian form, requires quite modern and complicated
group-theoretic tools which are connected with a construction  of
different representations of algebras and superalgebras underlying
above theories. Whereas for Lie (super)algebra case relevant for
HS fields on flat spaces the finding and structure of mentioned
objects like Verma modules and generalized Verma modules
\cite{Dixmier}, \cite{Dobrev} are rather understandable, an
analogous situation with non-linear algebraic and superalgebraic
structures which corresponds to
 HS fields on AdS spaces has not been classified to
present  with except for the case of totally-symmetric bosonic
\cite{BurdikNavratilPasnev}--\cite{BKL} and fermionic
\cite{adsfermBKR}, \cite{0905.2705} HS fields. The paper propose
the results of regular method for Verma module constructing and
 Fock space realization for quadratic algebras, whose negative
(equivalently positive) root vectors in Cartan-like triangular
decomposition are entangled due to presence  of the special
parameter $r$ being by the inverse square of AdS-radius vanishing
in the flat space limit.
The obtained objects permit to find
 Lagrangian formulations (LFs) for free integer  HS fields on
$d$-dimensional $AdS$-space  with $Y(s_1,...,s_k)$ in Fronsdal
\cite{Fronsdal} metric-like formalism within  BFV-BRST procedure
\cite{BFV}, \cite{Henneaux} (known at present as BRST construction) as a starting
point for an interacting HS field theory in the framework of
conventional Quantum Field Theory. The application of the BRST
construction to free HS field theory on AdS spaces consists in 4
steps and presents a solution of the \textit{problem inverse} to
that of the method \cite{BFV} (as in the case of string field
theory \cite{SFT})and reflects one more side  of the BV--BFV
duality concept \cite{GrigorievDamgaard}--\cite{GMR1}. First, the conditions
that determine the representations with  given mass and spin are
regarded as a topological (i.e. without Hamiltonian) gauge system
of mixed-class operator constraints $o_I$, $I=1,2,...$, in an
auxiliary Fock space $\mathcal{H}$. Second, the whole system of
$o_I$ which  form quadratic commutator algebra should be
additively  converted (see \cite{conversion}, \cite{conversion1} for the development
of conversion methods) in deformed (in power of parameter $r$)
algebra of  $O_I$: $O_I=o_I+o'_I$ determined on wider Fock space,
$\mathcal{H}\bigotimes \mathcal{H}'$ with first-class constraints
$O_\alpha$, $O_\alpha \subset O_I$. Third, the,
  Hermitian nilpotent BFV-BRST operator $Q'$ for non-linear algebra of
 converted operators $O_I$ which  contains the  BFV-BRST operator $Q$ for only
 subsystem of $O_\alpha$ should be found. Fourth, the Lagrangian
 $\mathcal{L}$ for given HS field  through a scalar product
 $\langle \ | \ \rangle$ in total Hilbert space $\mathcal{H}_{tot}$ like
$\mathcal{L} \sim \langle \phi |Q |\phi \rangle$, to be invariant
with respect to (reducible) gauge transformations $\delta|\phi
\rangle = Q |\Lambda \rangle$ with $|\phi \rangle$ containing
initial and auxiliary HS fields  is constructed in such a way that
the corresponding equations of motion reproduce the initial
constraints.

The application of above algorithm  for bosonic \cite{flatbos}, \cite{BuchbinderKrykhtin} and
fermionic \cite{flatferm}--\cite{flatfermmix}
 HS fields on flat spaces  did not meet the
problems in rather complicated second and third steps due to
linear Lie structure of   initial constraints $o_I$ algebra,
$[o_I, o_J\} = f_{IJ}^K o_K$ for structure constants $f_{IJ}^K$.
Indeed, for the same algebra of additional parts $o'_I$ it is
sufficient to use the construction of Verma module (VM) for
integer HS symmetry algebra $sp(2k)$ \cite{BuchbinderReshetnyak}
which is in one-to-one correspondence with  Lorentz  $so(1,d-1)$
algebra unitary irreducible representations subject to
$Y(s_1,...,s_k)$, $k\leq \left[\frac{d}{2}\right]$ due to Howe
duality \cite{Howe1}. Then an oscillator realization of the
symplectic algebra $sp(2k)$ in Fock space $\mathcal{H}'$
\cite{BuchbinderReshetnyak} represents a polynomial form as
compared to totally-symmetric HS fields on AdS space
\cite{symint-adsmassless}--\cite{0905.2705}
 where, (super)algebras of $o_I$ are
non-linear, not coinciding with  ones for $o'_I$. The  problem of the same complication arises in
constructing of BFV-BRST operator $Q'$ for (super)algebra of
converted constraints $O_I$ which in transiting to AdS-space does
not have the  form to be quadratic in ghost coordinates
$\mathcal{C}$ and requires the full study of  algebraic relations
starting from  Jacobi identities resolution (see \cite{0812.2329}
for details of finding BRST operator $Q'$ in question
     and ones for classical quadratic algebras \cite{bl}).

To be complete, note the details of Lagrangian description of
mixed-symmetry  HS tensors on  (A)dS backgrounds  were studied in
"frame-like" formulation in \cite{Alkalaev}--\cite{Alkalaev2} whereas the LF
for the
 mixed-symmetry bosonic fields
with off-shell traceless constraints  in the case of
(anti)-de-Sitter case are recently known for the
  Young tableaux with two rows
 \cite{Zinoviev1}, \cite{Zinoviev2}. The
 aspects of $SO(N)$ spinning particles dynamics after applying
 of Dirac
quantization procedure to  particle's first-class constraints
system which produces the dynamics of HS fields on constant
curvature spaces were studied in \cite{Latini}. At last,
 the various aspects of mixed-symmetry HS
fields  Lagrangian dynamics on Minkowski space  were
  discussed in \cite{Franciamix}, \cite{Franciamix1} and recently  for interacting
  mixed-symmetry HS fields on AdS-spaces  in \cite{interlow}, \cite{interlow1}.

Present paper is devoted to the  solution of the following
problems:
\begin{enumerate}
\item derivation of  HS symmetry algebra   for
  bosonic  HS fields in $d$-dimensional  AdS space
subject to arbitrary YT $Y(s_1,...,s_k)$;
    \item development of a method of Verma module construction
    for a non-linear HS symmetry algebra
      for    YT with two rows $Y(s_1,s_2)$
      and to the oscillator realization
    for given non-linear algebra as  formal power series
    in  creation and annihilation operators
     of corresponding Heisenberg algebra;
    \item construction of  unconstrained  LF for free bosonic
    HS fields on AdS-space with $Y(s_1,s_2)$.
\end{enumerate}

The paper is organized as follows. In Section~\ref{HSfieldsAdS},
we examine the bosonic HS fields that includes, first,  the
derivation of  HS symmetry algebra $\mathcal{A}(Y(k), AdS_d)$ for
HS fields subject to Young tableaux with arbitrary number of rows
$k$ in Subsection~\ref{HSalg}, second, an auxiliary proposition that
permits one to find the forms of deformation of general commutator
polynomial  algebras under additive conversion procedure in
Subsection~\ref{auxtheorem}. In Section~\ref{HSauxY2b}  we derive
explicitly HS symmetry algebra of additional parts for
HS fields with $2$ families  of indices, formulate and solve the problem of
Verma module construction for
algebra $\mathcal{A}'(Y(2), AdS_d)$  of additional parts $o'_I$ in Subsection~\ref{VMboconic}.
 We find  Fock space realization for the algebra
$\mathcal{A}'(Y(2), AdS_d)$  in Subsection~\ref{oscVMY2b}. In
Section~\ref{Lagrform}, we derive an explicit form for
non-linear algebra $\mathcal{A}_c(Y(2), AdS_d)$ of converted
operators $O_I$, and  present for it
an expression for BFV--BRST operator  in  Subsection\ref{BRSTk2}.
  and
 develop  the unconstrained
Lagrangian formulation for bosonic HS fields with two-rows Young
tableaux $Y(s_1,s_2)$ in Subsection~\ref{prLagrform}.
 In Conclusion, we summarize the results of the
work and discuss some open problems. Finally, in \ref{proof} we
prove the proposition on additive conversion for polynomial algebras.
\section{HS fields in AdS spaces with integer
spin} \label{HSfieldsAdS}
In the section  we  derive  numbers of special HS symmetry
non-linear algebras which encode mixed-symmetry tensor fields as
the elements of AdS group irreducible representations with
generalized spin $\mathbf{s}=(s_1,\ldots , s_k)$ and mass $m$ on
Ads${}_d$-space-time. We consider the problem of Verma module
construction for one of them and solve it explicitly  for
non-linear algebra with two-rows Young tableaux. The construction of the Fock
space representation for the non-linear algebra with found Verma
module finishes the solution of the problem there.
\subsection{HS symmetry algebra $\mathcal{A}(Y(k), AdS_d)$ for
mixed-symmetry tensor fields with  $Y(s_1,...,s_k)$}\label{HSalg}
A massive generalized integer spin $\mathbf{s}=(s_1,...,s_k)$,
($s_1 \geq s_2\geq ... \geq s_k>0$, $k \leq [d/2]$), AdS group
irreducible representation   in an AdS${}_d$ space is realized in
a space of mixed-symmetry tensors,
$\Phi_{(\mu^1)_{s_1},(\mu^2)_{s_2},...,(\mu^k)_{s_k}}
\hspace{-0.2em}\equiv \hspace{-0.2em}
\Phi_{\mu^1_1\ldots\mu^1_{s_1},\mu^2_1\ldots\mu^2_{s_2},...,
\mu^k_1\ldots \mu^k_{s_k}}(x)$ to be corresponding to a Young
tableaux
\begin{equation}\label{Young k2}
\Phi_{(\mu^1)_{s_1},(\mu^2)_{s_2},...,(\mu^k)_{s_k}}
\hspace{-0.3em}\longleftrightarrow \hspace{-0.3em}
\begin{array}{|c|c|c c c|c|c|c|c|c| c|}\hline
  \!\mu^1_1 \!&\! \mu^1_2\! & \cdot \ & \cdot \ & \cdot \ & \cdot\  & \cdot\  & \cdot\ &
  \cdot\    &\!\! \mu^1_{s_1}\!\! \\
   \hline
    \! \mu^2_1\! &\! \mu^2_2\! & \cdot\
   & \cdot\ & \cdot  & \cdot &  \cdot & \!\!\mu^2_{s_2}\!\!   \\
  \cline{1-8} \cdot\ & \cdot\ & \cdot\
   & \cdot\ & \cdot  & \cdot &  \cdot & \cdot\!\!   \\
   \cline{1-8}
    \! \mu^k_1\! &\! \mu^k_2\! & \cdot\
   & \cdot\ & \cdot  & \cdot &   \!\!\mu^k_{s_k}\!\!   \\
   \cline{1-7}
\end{array}\ ,
\end{equation}
subject to the Klein-Gordon (\ref{Eq-0b}), divergentless
(\ref{Eq-1b}), traceless (\ref{Eq-2b}) and mixed-symmetry
equations (\ref{Eq-3b}) [for $\beta = (2;3;...;k+1)
\Longleftrightarrow (s_1>s_2; s_1 = s_2>s_3;...;s_1 =
s_2=...=s_k)$]  \cite{Metsaev}:
\begin{eqnarray}
\label{Eq-0b} &&\bigl[\nabla^2 +r[(s_1-\beta-1+ d)(s_1-\beta) -
\sum_{i=1}^ks_i]+m^2
\bigr]\Phi_{(\mu^1)_{s_1},(\mu^2)_{s_2},...,(\mu^k)_{s_k}}
 =0,\\
&&\nabla^{\mu^i_{l_i}}\Phi_{
(\mu^1)_{s_1},(\mu^2)_{s_2},...,(\mu^k)_{s_k}} =0, \quad
i,j=1,...,k;\, l_i,m_i=1,...,s_i\,, \label{Eq-1b}
\\
&& g^{\mu^i_{l_i}\mu^i_{m_i}}\Phi_{
(\mu^1)_{s_1},(\mu^2)_{s_2},...,(\mu^k)_{s_k}}=
g^{\mu^i_{l_i}\mu^j_{m_j}}\Phi_{
(\mu^1)_{s_1},(\mu^2)_{s_2},...,(\mu^k)_{s_k}} =0, \quad
 l_i<m_i,  \,,\label{Eq-2b}\\
&& \Phi_{
(\mu^1)_{s_1},...,\{(\mu^i)_{s_i}\underbrace{,...,\mu^j_{1}...}\mu^j_{l_j}\}...\mu^j_{s_j},...(\mu^k)_{s_k}}=0,\quad
i<j,\ 1\leq l_j\leq s_j, \label{Eq-3b}
\end{eqnarray}
where the brackets below denote that the indices  inside it do not
include in  symmetrization, i.e. the symmetrization concerns only
indices $(\mu^i)_{s_i}, \mu^j_{l_j} $ in
$\{(\mu^i)_{s_i}\underbrace{,...,\mu^j_{1}...}\mu^j_{l_j}\}$.

 To obtain HS symmetry algebra (of $o_I$) for a
 description of all integer spin HS fields, we in a standard
 manner
 introduce a Fock
space $\mathcal{H}$, generated by $k$ pairs of bosonic creation
$a^i_{\mu^i}(x)$ and annihilation $a^{j+}_{\nu^j}(x)$ operators,
$i,j =1,...,k, \mu^i,\nu^j  =0,1...,d-1$\footnote{such choice of
the oscillators corresponds to the case of symmetric basis,
whereas there exists another realization of auxiliary Fock space
generated by the fermionic oscillators (antisymmetric basis)
$\hat{a}^m_{\mu^m}(x)$, $\hat{a}^{\hat{n}+}_{\nu^n}(x)$ with
anticommutation relations, $\{\hat{a}^m_{\mu^m},
\hat{a}_{\nu^n}^{n+}\}=-g_{\mu^m\nu^m}\delta^{mn}$,   for $m, n =
1,..., s_1$, and develop the  procedure below following to the
lines of Ref. \cite{brst1} for totally antisymmetric tensors for
$s_1=s_2=...=s_k=1$.}
\begin{eqnarray}\label{comrels}
[a^i_{\mu^i}, a_{\nu^j}^{j+}]=-g_{\mu^i\nu^j}\delta^{ij}\,, \qquad
\delta^{ij} = diag(1,1,\ldots 1)\,,
\end{eqnarray}
 and a set of constraints for an arbitrary string-like vector
$|\Phi\rangle \in \mathcal{H}$ which we call as the basic vector,
\begin{eqnarray}
\label{PhysState}  \hspace{-2ex}&& \hspace{-2ex} |\Phi\rangle  =
\sum_{s_1=0}^{\infty}\sum_{s_2=0}^{s_1}\cdots\sum_{s_k=0}^{s_{k-1}}
\Phi_{(\mu^1)_{s_1},(\mu^2)_{s_2},...,(\mu^k)_{s_k}}(x)\,
\prod_{i=1}^k\prod_{l_i=1}^{s_i} a^{+\mu^i_{l_i}}_i|0\rangle,\\
\label{l0} \hspace{-2ex}&& \hspace{-3ex} {\tilde{l}}_0|\Phi\rangle
= \bigl(l_0+ \tilde{m}^2_b + r \bigl((g_0^1-2\beta-2)g_0^1 -
 \sum^k_{i=2}g_0^i \bigr)\bigr)|\Phi\rangle=0 , \quad l_0 = [D^2 -
r\textstyle\frac{d(d-2(k+1))}{4}],\\
\label{lilijt} \hspace{-2ex} && \hspace{-2ex} \bigl({l}^i, l^{ij},
t^{i_1j_1} \bigr)|\Phi\rangle  = \bigl(-i a^i_\mu D^\mu,
\textstyle\frac{1}{2}a^{i}_\mu a^{j\mu}, a^{i_1+}_\mu
a^{j_1\mu}\bigr) |\Phi\rangle=0,\quad i\leq j;\, i_1 < j_1,
\end{eqnarray}
with number particles operators, central charge, covariant
derivative in $\mathcal{H}$ respectively,
\begin{eqnarray}
\label{numpart} && g_0^i = -\frac{1}{2}\{a^{i+}_\mu,
a^{\mu{}i}\},\qquad \qquad \qquad\   \tilde{m}^2_b = {m}^2 + r\beta(\beta+1),\\
&&  D_\mu =
\partial_\mu - \omega_\mu^{ab}(x)\Bigl(\sum_{i}a_{i{}a}^+a_{i{}b}
\Bigr), \qquad a^{\mu(+)}_i(x) =
e^\mu_a(x)a^{a(+)}_i\footnotemark,
\end{eqnarray}
\footnotetext{operators $a^{a_i}_i$, $a^{b_j+}_j$ satisfy to usual
for $R^{1,d-1}$-space commutation relations
$[a^{a_i}_i,a^{b_j+}_j]=-\eta^{a_ib_i}\delta_{ij}$ for
$\eta^{ab}=\mathrm{diag}(+,-,\ldots,-)$}where $e^\mu_a$,
 $\omega_\mu^{ab}$ are vielbein and spin connection for
tangent indices $a,b = 0,1...,d-1$. Operator $D_\mu$  is equivalent in its action in
$\mathcal{H}$ to the covariant derivative $\nabla_{\mu}$ [with
d'Alambertian $D^2 = (D_a + \omega^{b}{}_{ba} )D^a$]. The set of
$k(k+1)$ primary constraints (\ref{l0}), (\ref{lilijt}) with
$\{o_\alpha\}$ = $\bigl\{{\tilde{l}}_0, {l}^i, l^{ij}, t^{i_1j_1}
\bigr\}$ are equivalent to Eqs. (\ref{Eq-0b})--(\ref{Eq-3b}) for
all admissible values of spins and for the field
$\Phi_{(\mu^1)_{s_1},(\mu^2)_{s_2},...,(\mu^k)_{s_k}}$ with fixed
spin $\mathbf{s}=(s_1,s_2,\ldots , s_k)$
 if in addition to  Eqs. (\ref{l0}), (\ref{lilijt}) we add $k$
  constraints  with  $g_0^i$,
\begin{eqnarray}\label{g0iphys}
g_0^i|\Phi\rangle =\textstyle(s_i+\frac{d}{2}) |\Phi\rangle.
\end{eqnarray}

The requirement of closedness of the algebra with $o_\alpha$ with
respect to $[\ ,\ ]$-multiplication leads to   enlargement of
$o_\alpha$ by adding the operators $g_0^i$ and hermitian
conjugated  operators $o_\alpha^+$,
\begin{equation}\label{lilijt+}
 \bigl({l}^{i+},\
l^{ij+},\ t^{i_1j_1+} \bigr)  = \bigl(-i a^{i+}_\mu D^\mu,\
\textstyle\frac{1}{2}a^{i+}_\mu a^{j\mu+},\ a^{i_1}_\mu
a^{j_1\mu+}\bigr) ,\ i\leq j;\ i_1 < j_1,
\end{equation}
 with respect to  scalar product on $\mathcal{H}$,
\begin{eqnarray}
\label{sproduct} \langle{\Psi}|\Phi\rangle & =  & \int d^dx
\sqrt{|g|}\sum_{i=1}^k\sum_{s_i=0}^{s_{i-1}}
         \sum_{j=1}^k\sum_{p_j=0}^{s_{j-1}}
\langle 0|\prod_{j=1}^k\prod_{m_j=1}^{p_j}
a^{\nu^j_{m_j}}_j\Psi^*_{(\nu^1)_{p_1},(\nu^2)_{p_2},...,(\nu^k)_{p_k}}(x)\nonumber\\
&& \times
\Phi_{(\mu^1)_{s_1},(\mu^2)_{s_2},...,(\mu^k)_{s_k}}(x)\,
\prod_{i=1}^k\prod_{l_i=1}^{s_i}
a^{+\mu^i_{l_i}}_i|0\rangle,\texttt{ for } s_{-1}, p_{-1} =
\infty.
\end{eqnarray}
This fact  will guarantee the Hermiticity of corresponding
BFV-BRST operator with taken into account of self-conjugated
operators, $(l_0^+,\ {g_0^i}^+) = (l_0,\ {g_0^i})$ (therefore the
reality of Lagrangian $\mathcal{L}$) for the system  of all
operators $\{o_\alpha, o_\alpha^+, g_0^i\}$. We call the algebra
of these operators the \emph{integer higher-spin symmetry algebra
in AdS space with a Young tableaux having $k$ rows}\footnote{one
should not confuse the term "\emph{higher-spin symmetry algebra}"
using here for free HS formulation  with the algebraic structure
known as "\emph{higher-spin algebra}" (see, e.g.
Ref.\cite{Vasiliev_inter}) arising to describe the HS
interactions} and denote it as  $\mathcal{A}(Y(k), AdS_d))$.

  The maximal Lie subalgebra of operators $l^{ij},
t^{i_1j_1}, g_0^i, l^{ij+},\ t^{i_1j_1+}$ is isomorphic to
symplectic algebra $sp(2k)$ (see, \cite{BuchbinderReshetnyak} for
details and we will  refer on it later as $sp(2k)$) whereas the
only nontrivial quadratic commutators in $\mathcal{A}(Y(k),
AdS_d))$ are due to operators with $D_{\mu}$: $l^i, \tilde{l}_0,
l^{i+}$. For the aim of LF construction it is enough to have a
simpler, without central charge $\tilde{m}^2_b$ (so called
\textit{modified} HS symmetry algebra $\mathcal{A}_{m}(Y(k),
AdS_d))$, with operator $l_0$ (\ref{l0}) instead of
${\tilde{l}}_0$, so that
 AdS-mass term, $\tilde{m}^2_b + r
\bigl((g_0^1-2\beta-2)g_0^1 -
 \sum_{i=2}^kg_0^i \bigr)$, will be restored as usual later within
 conversion procedure
 and  properly construction of LF.

Algebra $\mathcal{A}_{m}(Y(k), AdS_d))$  of the operators $o_I$
from the Hamiltonian analysis of the dynamical systems viewpont
contains $1$ first-class constraint $l_0$,   $2k$ differential
$l_i, l_i^+ $ and $2k^2$ algebraic $t^{i_1j_1}, t^+_{i_1j_1},
l^{ij}, l_{ij}^+$ second-class constraints $o_{{a}}$ and operators
$g_0^i$, composing an invertible matrix $\Delta_{ab}(g_0^i)$ for
topological gauge system  because of
\begin{eqnarray}
[o_a,\; o_b] = f^c_{ab} o_c + f^{cd}_{ab} o_co_d +
\Delta_{ab}(g_0^i),\ [o_a,\;l_o] = f^c_{a{}[l_0]}o_c +
f^{cd}_{a[l_0]} o_co_d,  . \label{inconstraintsd}
\end{eqnarray}
Here $f^c_{ab},  f^{cd}_{ab}, f^c_{a{}[l_0]}, f^{cd}_{a[l_0]},
\Delta_{ab}$ are  antisymmetric with respect to permutations of
lower indices constant quantities and the operators
$\Delta_{ab}(g_0^i)$ form the non-vanishing $2k(k+1)\times
2k(k+1)$ matrix $\|\Delta_{ab}\|$ in the Fock space $\mathcal{H}$
on the surface $\Sigma \subset \mathcal{H}$:
$\|\Delta_{ab}\|_{|\Sigma} \ne 0 $, which is determined by the
equations, $o_\alpha|\Phi\rangle = 0$. The set of $o_I$  satisfies
the non-linear relations  (additional to ones for $sp(2k)$) given
by the multiplication table~\ref{table}.
\hspace{-1ex}{\begin{table}[t]\caption{HS symmetry non-linear algebra  $\mathcal{A}_{m}(Y(k),
AdS_d)$.\label{table} } {{
\begin{center}
\begin{tabular}{||c||c|c|c|c|c|c|c||}\hline\hline
$\hspace{-0.2em}[\; \downarrow, \rightarrow
\}\hspace{-0.5em}$\hspace{-0.7em}&
 $t^{i_1j_1}$ & $t^+_{i_1j_1}$ &
$l_0$ & $l^i$ &$l^{i{}+}$ & $l^{i_1j_1}$
&$l^{i_1j_1{}+}$  \\
\hline\hline $t^{i_2j_2}$
    & $A^{i_2j_2, i_1j_1}$ & $B^{i_2j_2}{}_{i_1j_1}$&$0$
   & \hspace{-0.3em}
    $\hspace{-0.2em}l^{j_2}\delta^{i_2i}$\hspace{-0.5em} &
    \hspace{-0.3em}
    $-l^{i_2+}\delta^{j_2 i}$\hspace{-0.3em}
    &\hspace{-0.7em} $\hspace{-0.7em}l^{\{j_1j_2}\delta^{i_1\}i_2}
    \hspace{-0.9em}$ \hspace{-1.2em}& \hspace{-1.2em}$\hspace{-0.9em}
    -l^{i_2\{i_1+}\delta^{j_1\}j_2}\hspace{-0.9em}$\hspace{-1.2em} \\
\hline $t^+_{i_2j_2}$
    & $-B^{i_1j_1}{}_{i_2j_2}$ & $A^+_{i_1j_1, i_2j_2}$&$0$
   & \hspace{-0.3em}
    $\hspace{-0.2em} l_{i_2}\delta^{i}_{j_2}$\hspace{-0.5em} &
    \hspace{-0.3em}
    $-l^+_{j_2}\delta^{i}_{i_2}$\hspace{-0.3em}
    & $l_{i_2}{}^{\{j_1}\delta^{i_1\}}_{j_2}$ & $-l_{j_2}{}^{\{j_1+}
    \delta^{i_1\}}_{i_2}$ \\
\hline $l_0$
    & $0$ & $0$&$0$
   & \hspace{-0.3em}
    $\hspace{-0.2em}-r{\mathcal{K}}^{i+}_1$\hspace{-0.5em} & \hspace{-0.3em}
    $r{\mathcal{K}}^{i}_1$\hspace{-0.3em}
    & $0$ & $0$  \\
\hline $l^j$
   & \hspace{-0.5em}$- l^{j_1}\delta^{i_1j}$ \hspace{-0.5em} &
   \hspace{-0.5em}$
   -l^{i_1}\delta^{j_1j}$ \hspace{-0.9em}  &\hspace{-0.3em}
    $\hspace{-0.2em}r{\mathcal{K}}^{j+}_1$\hspace{-0.5em} \hspace{-0.3em}&${W}^{ji}_b$ \hspace{-0.3em} & \hspace{-0.3em}
   ${X}^{ji}_b$\hspace{-0.3em}
    & $0$ & \hspace{-0.5em}$- \textstyle\frac{1}{2}l^{\{i_1+}\delta^{j_1\}j}$
    \hspace{-0.9em}  \\
\hline $l^{j+}$ & \hspace{-0.5em}$l^{i_1+}
   \delta^{j_1j}$\hspace{-0.7em} & \hspace{-0.7em}$l^{j_1+}\delta^{i_1j}$ \hspace{-1.0em} &\hspace{-0.3em}
    $\hspace{-0.2em}-r{\mathcal{K}}^{j}_1$\hspace{-0.5em}
   &\hspace{-0.3em}
   $-{X}^{ij}_b$\hspace{-0.3em}
   &\hspace{-0.5em} $- {W}^{ji+}_b$\hspace{-0.5em}
    &\hspace{-0.7em} $ \textstyle\frac{1}{2}l^{\{i_1}\delta^{j_1\}j}
    $\hspace{-0.7em} & $0$  \\
\hline $l^{i_2j_2}$
    & \hspace{-0.2em}$\hspace{-0.2em}-l^{j_1\{j_2}\delta^{i_2\}i_1}
    \hspace{-0.5em}$
    \hspace{-0.5em} &\hspace{-0.2em} $\hspace{-0.4em}
    -l^{i_1\{i_2}\delta^{j_2\}j_1}\hspace{-0.3em}$\hspace{-0.3em}&$0$
   & \hspace{-0.3em}
    $0$\hspace{-0.5em} & \hspace{-0.3em}
    $ \hspace{-0.7em}-\textstyle\frac{1}{2}l^{\{i_2}\delta^{j_2\}i}
    \hspace{-0.5em}$\hspace{-0.3em}
    & $0$ & \hspace{-0.7em}$\hspace{-0.3em}
L^{i_2j_2,i_1j_1}\hspace{-0.3em}$\hspace{-0.7em}  \\
\hline $l_{i_2j_2}^+$
    & $ l^{i_1+}_{\, \{i_2}\delta^{j_1}_{j_2\}}$  & $ l^{+}_{j_1\{j_2}
    \delta_{i_2\}i_1}$ & $0$
   & \hspace{-0.3em}
    $\hspace{-0.2em} \textstyle\frac{1}{2}l^{+}_{\{i_2}\delta^{i}_{j_2\}}$\hspace{-0.5em} & \hspace{-0.3em}
    $0$\hspace{-0.3em}
    & $-L^{i_1j_1,i_2j_2}$ & $0$  \\
\hline\hline $g^j_0$
    & $-F^{i_1j_1,j}$ & $F^{i_1j_1,j+}$&$0$
   & \hspace{-0.3em}
    $\hspace{-0.2em}-l^i\delta^{ij}$\hspace{-0.5em} & \hspace{-0.3em}
    $l^{i+}\delta^{ij}$\hspace{-0.3em}
    & \hspace{-0.7em}$\hspace{-0.7em}  -l^{j\{i_1}\delta^{j_1\}j}\hspace{-0.7em}$\hspace{-0.7em} & $ l^{j\{i_1+}\delta^{j_1\}j}$ \\
   \hline\hline
\end{tabular}
\end{center}}}\end{table}

First note that,  in the table~\ref{table}   we did not include
the columns with $[\ ,\ \}$-products of all $o_I$ with $g_0^j$
which may be obtained from the rows with $g_0^j$  as follows: $[b
,o_I\}^+$ = $-[o^+_I ,b \}$, with account of closedness of HS
algebra with respect to Hermitian conjugation. Second the
operators $t^{i_2j_2}, t^+_{i_2j_2}$ satisfy by the definition the
properties
\begin{equation} \label{thetasymb} (t^{i_2j_2}, t^+_{i_2j_2}) \equiv
(t^{i_2j_2},t^+_{i_2j_2})\theta^{j_2i_2}, \ \theta^{j_2i_2} =
(1,0)\verb" for " (j_2> i_2, j_2\leq i_2)
\end{equation}
with Heaviside $\theta$-symbol\footnote{there are no summation
with respect to the indices $i_2, j_2$ in the
Eqs.(\ref{thetasymb}), the figure brackets for the indices $i_1$,
$i_2$ in the quantity $A^{\{i_1}B^{i_2\}i_3}\theta^{i_3i_2\}}$
mean the symmetrization
 $A^{\{i_1}B^{i_2\}i_3}\theta^{i_3i_2\}}$ =
$A^{i_1}B^{i_2i_3}\theta^{i_3i_2}+
A^{i_2}B^{i_1i_3}\theta^{i_3i_1}$ as well as these indices are
raising and lowering by means of Euclidian metric tensors
$\delta^{ij}$, $\delta_{ij}$, $\delta^{i}_{j}$} $\theta^{ji}$.
Third, the products $B^{i_2j_2}_{i_1j_1}, A^{i_2j_2, i_1j_1},
F^{i_1j_1,i}, L^{i_2j_2,i_1j_1}$ are determined by the explicit
relations,
\begin{eqnarray}
  {}B^{i_2j_2}{}_{i_1j_1} &=&
  (g_0^{i_2}-g_0^{j_2})\delta^{i_2}_{i_1}\delta^{j_2}_{j_1} +
  (t_{j_1}{}^{j_2}\theta^{j_2}{}_{j_1} + t^{j_2}{}^+_{j_1}\theta_{
  j_1}{}^{j_2})\delta^{i_2}_{i_1}
  -(t^+_{i_1}{}^{i_2}\theta^{i_2}{}_{i_1} + t^{i_2}{}_{i_1}\theta_{i_1}{}^{
  i_2})
  \delta^{j_2}_{j_1}
\,,\label{Bijkl}
\\
   A^{i_2j_2, i_1j_1} &=&  t^{i_1j_2}\delta^{i_2j_1}-
  t^{i_2j_1}\delta^{i_1j_2}  , \qquad    F^{i_2j_2,i} \ = \
   t^{i_2j_2}(\delta^{j_2i}-\delta^{i_2i}),\label{Fijk} \\
  L^{i_2j_2,i_1j_1} &=&   \textstyle\frac{1}{4}\Bigl\{\delta^{i_2i_1}
\delta^{j_2j_1}\Bigl[2g_0^{i_2}\delta^{i_2j_2} + g_0^{i_2} +
g_0^{j_2}\Bigr]  - \delta^{j_2\{i_1}\Bigl[t^{j_1\}i_2}\theta^{i_2j_1\}} +t^{i_2j_1\}+}\theta^{j_1\}i_2}\Bigr] \nonumber \\
&& - \delta^{i_2\{i_1}\Bigl[t^{j_1\}j_2}\theta^{j_2j_1\}}
+t^{j_2j_1\}+}\theta^{j_1\}j_2}\Bigr] \Bigr\}
 \,.\label{Lklij}
\end{eqnarray}
They obeys  the obvious additional properties of antisymmetry and
Hermitian conjugation,
\begin{eqnarray}
  &&\hspace{-1em} A^{i_2j_2,  i_1j_1} = -A^{ i_1j_1, i_2j_2}
  ,\qquad  A^+_{i_1j_1,  i_2j_2}=(A_{i_1j_1,  i_2j_2})^+ = t^+_{i_2j_1}\delta_{j_2i_1}
   -   t^+_{i_1j_2}\delta_{i_2j_1},\\
  &&\hspace{-1em} ({L^{i_2j_2,i_1j_1}})^+ =  L^{i_1j_1, i_2j_2} ,\qquad  {F^{i_2j_2,i+}}
  =(F^{i_2j_2,i})^+= t^{i_2j_2+}(\delta^{j_2i}-\delta^{i_2i}),\\
\label{Bijk+}
  && \hspace{-1em} {B^{i_2j_2}{}_{i_1j_1}}^+ = (g_0^{i_2}-g_0^{j_2})\delta^{i_2}_{i_1}
   \delta^{j_2}_{j_1} +
  (t^+_{j_1}{}^{j_2}\theta^{j_2}{}_{j_1} + t^{j_2}{}_{j_1}
  \theta_{j_1}{}^{j_2})\delta^{i_2}_{i_1}
  -(t_{i_1}{}^{i_2}\theta^{i_2}{}_{i_1} +
  t^{i_2+}{}_{i_1}\theta_{i_1}{}^{i_2})\delta^{j_2}_{j_1}.
\end{eqnarray}
Fourth, the independent quantities ${\mathcal{K}}^{k}_1,
{W}^{ki}_b, X^{ki}_b$ in the table~\ref{table} are quadratic in
$o_I$,
\begin{eqnarray}
{} {W}^{ij}_b & = & [l^i,l^j]= 2r\Bigl[(g_0^j-g_0^i)l^{ij} -
\sum_{m} (t^{m[j}\theta^{[jm}+
t^{[jm+}\theta^{m[j})l^{i]m}\Bigr],
 \label{lilj}\\
{\mathcal{K}}^{k}_1& =
&r^{-1}[l_0,l^{k+}]=\Bigl(4\sum_{i}l^{ki+}l^i +l^{k+}(2g_0^k-1)
-2(\sum_{i}l^{i+}t^{+ik}\theta^{ki}
+l^{i+}t^{ki}\theta^{ik})\Bigr)  \label{l'0li+} ,
\\
{} {X}^{ij}_b
  & =
 &\bigl\{{{l}}_{0}+ r\bigl( K^{0i}_0  +
\sum_{l=i+1}^k\mathcal{K}^{il}_0 +
\sum_{l=1}^{i-1}\mathcal{K}^{li}_0 \bigr)\bigr\}\delta^{ij}
      \label{lilj+b}\\
      && -
r\bigl[4\sum\nolimits_{l} l^{ jl+}l^{ li} -\sum_{l=1}^{j-1} t^{
lj+}t^{ li} -\sum_{l=i+1}^{k} t^{ il+}t^{jl}-\sum_{l=j+1}^{i-1}
t^{ li}t^{ jl} + \textstyle(g_0^{ j}+g_0^i
  -j-{1})t_{ji}\bigr]\theta^{ij}\nonumber\\
&& - r\bigl[4\sum\nolimits_{l} l^{ jl+}l^{ li} -\sum_{l=1}^{i-1}
t^{ lj+}t^{ li} -\sum_{l=j+1}^{k} t^{
il+}t^{jl}-\sum_{l=i+1}^{j-1} t^{ il+}t^{ lj+}+ t^+_{ij}(g_0^{
j}+g_0^{ i}-i
  -{1}
)\bigr]\theta^{ji}. \nonumber
\end{eqnarray}
In writing (\ref{lilj+b}) we have used the quantities  $K^{0i}_0,
i,j=1,...,k$, $\mathcal{K}^{ij}_0$ composing a Casimir operator
$\mathcal{K}_0(k)$ for $sp(2k)$ algebra
\begin{equation}\label{Casimirsb}
\mathcal{K}_0(k) =
   \sum_{i}K_0^{0i} +
    2\sum_{i,j}\mathcal{K}_0^{ij}\theta^{ji} =
    \sum_{i}\bigl((g_0^i)^2-2g_0^i -4l^{ii+}l^{ii}\bigr)+2
    \bigl(\sum_{i=1}^k\sum_{j=i+1}^k(t^{ +}_{ij}t^{ij}-
    4l^{ +}_{ij}
      l^{ {ij}} -
    g_0^{ j}) \bigr).
\end{equation}
Algebra $\mathcal{A}_m(Y(k), AdS_d)$ maybe considered as
non-linear deformation in power of $r$ of the integer HS symmetry
algebra in Minkowski space  $\mathcal{A}(Y(k), {R}^{1,d-1})$
\cite{BuchbinderReshetnyak} as follows,
\begin{equation}\label{identalg}
  \mathcal{A}_m(Y(k), AdS_d) =  \mathcal{A}(Y(k),
  {R}^{1,d-1})(r)
  = \left(T^k \oplus T^{k*}\oplus l_0\right)(r) + \hspace{-1em} \supset
  sp(2k),
\end{equation}
for $k$-dimensional commutative (on ${R}^{1,d-1}$) algebra $T^k =
\{l_i\}$, its dual $T^{k*}=\{l^{i+}\}$ and represents the
semidirect sum of the symplectic algebra $sp(2k)$ [as an algebra
of internal derivations of $(T^k \oplus T^{k*})]$. For HS fields
with single spin $s_1$ (for $k=1$) and with two-component spin
$(s_1,s_2)$ (for $k=2$) the algebra $\mathcal{A}_m(Y(k), AdS_d)$
coincides respectively with known HS symmetry  algebras on AdS
spaces given in \cite{BKL} and in \cite{0812.2329}.

Now, we should to use the results of an special proposition in order
to proceed to the conversion procedure of the algebra of $o_I$  to get the algebra of $O_I$ with only first-class
constraints.

\subsection{On additive conversion for
polynomial algebras}\label{auxtheorem}

In this subsection, to solve the problem of  additive conversion
of non-linear algebras  with a subset of 2nd class constraints, we
need to use some important statement which based on the following
(see Ref.\cite{0905.2705} for
 detailed description)

\noindent \textbf{\emph{Definition:}} \emph{A non-linear
commutator algebra $\mathcal{A}$ of basis elements $o_I$, $I \in
\Delta$ (with  $\Delta$ to be finite or infinite set of indices)
is called a \textbf{polynomial
 algebra of order $n$}, $n \in \mathbf{N}$, if a set of
$\{{o}_I\}$ is subject to $n$-th order polynomial commutator
relations:}
\begin{eqnarray}\label{polynom} && [o_I, o_J]  =
        F^{K}_{IJ}(o)o_{K},\
        F^{K}_{IJ} = f^{(1)K}_{IJ} + \sum_{n=2}^\infty
        f^{(n)K_1...K_{n-1}K}_{IJ}\prod_{i=1}^{n-1} o_{K_i}\,,
 \nonumber \\
{}&&\qquad f^{(n)K_1...K_n}_{IJ} \neq 0, \texttt{ and }
f^{(k)K_1...K_nK_{n+1}...K_k}_{IJ}=0,\, k>n,
\end{eqnarray}
\emph{with structural coefficients $f^{(n)K_1...K_{n-1}K}_{IJ}$},
 to be  antisymmetric with respect
to permutations of lower indices, $f_{IJ}^{(n)K_1\cdots K_n} =
-f_{JI}^{(n)K_1\cdots K_n}$.

Now, we may to formulate the basic statement in the
subsection~\ref{auxtheorem} in the form of

\noindent  \textbf{\emph{Proposition:}} Let $\mathcal{A}$ is the
polynomial algebra of order $n$  of basis elements $o_I$
determined in Hilbert  space $\mathcal{H}$. Then, for a set
$\mathcal{A}'$ of  elements $o'_I$  given in a new Hilbert space
$\mathcal{H}'$ ($\mathcal{H} \bigcap \mathcal{H}' = \emptyset$)
and commuting with $o_I$, and   for a direct sum of sets
$\mathcal{A}_{c}= \mathcal{A} + \mathcal{A}'$ of the operators,
${O}_I$, ${O}_I = {o}_I + {o}'_I$ given in the tensor product
$\mathcal{H}\otimes \mathcal{H}'$ from the requirement to be in
involution relations,
\begin{equation}\label{involrel}
 [{O}_I,{O}_J] =
{F}_{IJ}^K({o}',{O}) {O}_K\,,
\end{equation} it follows the sets of $\{{o}'_I\}$, $\{{O}_I\}$
form respectively the polynomial commutator algebra $\mathcal{A}'$
of order $n$  and the non-linear commutator algebra
$\mathcal{A}_{c}$ with composition laws:
\begin{eqnarray} \label{addalg}
 && [\,o_I',o_J'] = f_{IJ}^{(1)K_1}{o}'_{K_1}+ \sum_{m=2}^n(-1)^{
  m-1}f_{IJ}^{(m)K_{l}\cdots
K_1}\prod_{s=1}^{m}{o}'_{K_s},\\ \label{conv-alg}
  && [\,{O}_I,{O}_J] \hspace{-0,3ex}=\hspace{-0,3ex}
  \Bigl(f_{IJ}^{(1)K} + \sum_{m=2}^{n}
  F^{(m){}K}_{IJ}({o}',{O})\Bigr){O}_{K}.
\end{eqnarray}
 The structural functions
$F^{(m){}k}_{IJ}(o',O)$ in (\ref{addalg}) are constructed with
respect to known from the Eqs. (\ref{polynom}) coefficients
$f^{(m)K_1...K_m}_{IJ} \equiv f^{K_1...K_m}_{IJ}$ as follows
\begin{eqnarray} \label{expfunc}
 \hspace{-1,5ex} && F^{(m){}K}_{IJ}
  \hspace{-0,3ex}=\hspace{-0,3ex}
f_{IJ}^{K_1\cdots K_{m}}\prod_{p=1}^{m-1}{O}_{K_p}
+\sum_{s=1}^{m-1}(-1)^{s} f_{ij}^{\widehat{K_s\cdots
K_1}\widehat{K_{s+1}\cdots K_{m}}} \prod_{p=1}^{s}{o}'_{K_{p}}
\prod_{l=s+1}^{m-1}{O}_{K_{l}},\verb" where "\nonumber \\
 \hspace{-1,5ex}&& f_{ij}^{\widehat{K_s\cdots K_1}\widehat{K_{s+1}\cdots K_{m}}} =
f_{ij}^{{K_s\cdots K_1}{K_{s+1}\cdots K_{m}}} + f_{ij}^{K_s\cdots
{K_{s+1}K_1}{K_{s+2}\cdots K_{m}}} +\cdots +
\nonumber \\
 \hspace{-1,5ex}{}&{}&  f_{ij}^{{K_{s+1}K_s\cdots K_1}{K_{s+2}\cdots
K_{m}}} +\Bigl(f_{ij}^{K_{s+1}K_s\cdots {K_{s+2}K_1}{K_{s+3}\cdots
K_{m}}} +\nonumber \\
 \hspace{-1,5ex}{}&{}&\cdots +
f_{ij}^{{K_{s+1}K_{s+2}K_s\cdots K_1}{K_{s+3}\cdots K_l}}\Bigr)+
\cdots + f_{ij}^{{K_{s+1}\cdots K_m}K_s\cdots
K_1},\label{sumcoeff}
\end{eqnarray}
%
where the sum in the Eq.(\ref{sumcoeff}) contains
$\frac{m!}{s!(m-s)!}$ terms with all the possible ways of the
arrangement the indices $(K_{s+1},..., K_{m})$ among the indices
$(K_{s},..., K_{1})$ in $f_{ij}^{{K_s\cdots K_1}{K_{s+1}\cdots
K_{m}}}$ without changing the separate orderings of the indices
$K_{s+1},..., K_{m}$ and $K_{s},..., K_{1}$\footnote{We do not
consider here the case of polynomial superalgebra for which the
proposition may be easily generalized with introducing
corresponding sign factors in the Eqs.
(\ref{addalg})--(\ref{sumcoeff}) with the same number of summands
to use it  for fermionic HS fields on AdS space}.

The validity of the proposition is verified  in the \ref{proof}.
Turning to the structure of  algebra $\mathcal{A}_c$ note that in
contrary to $\mathcal{A}$ and $\mathcal{A}'$  we  call it as
\emph{non-homogeneous polynomial algebra of order} $n$ due to form
of relations (\ref{expfunc}) (see footnote~13 in the \ref{proof}
for the comments). For Lie algebra case ($n=1$) the structures of
the algebras $\mathcal{A}$, $\mathcal{A}'$ and  $\mathcal{A}_c$
coincides as it then used, e.g. for the integer spin HS symmetry algebra
$\mathcal{A}(Y(k), {R}^{1,d-1})$ \cite{BuchbinderReshetnyak}. For
quadratic algebras (n=2) the algebraic relations for
$\mathcal{A}$, $\mathcal{A}^{\prime }$ and $\mathcal{A}_c^{}$ do not coincide
with each other due to  structural functions
$f^{(2)K_1K_{2}}_{IJ}$ presence that was firstly shown for the
algebra $\mathcal{A}(Y(1), AdS_d)$ for totally-symmetric HS
tensors on AdS space in \cite{BKL} and having the form,
\begin{eqnarray}
\hspace{-1em}&\hspace{-1em}&\hspace{-1em} [\,o_I,o_J]  =
f_{IJ}^{(1)K_1}{o}_{K_1}+
 f_{IJ}^{(2)K_{1}K_2}{o}_{K_1}{o}_{K_2},\quad [\,o'_I,o'_J] = f_{IJ}^{(1)K_1}{o}'_{K_1}-
f_{IJ}^{(2)K_{2}K_1}{o}'_{K_1}{o}'_{K_2}, \label{auxalg2}
 \\
 {}  \hspace{-1,7em}&\hspace{-1em}&\hspace{-1,7em} [\,{O}_I,{O}_J]  =
  \Bigl(f_{IJ}^{(1)K} +
  F^{(2){}K}_{IJ}({o}',{O})\Bigr){O}_{K},\quad
  F^{(2){}K}_{IJ}  = f_{IJ}^{(2)K_1K}{O}_{K_1} -
(f_{IJ}^{(2)K_1K} + f_{IJ}^{(2)KK_{1}}){o'}_{K_1}
\label{conv-alg2} .
\end{eqnarray}
Relations (\ref{auxalg2}), (\ref{conv-alg2}) are sufficient to
determine the form of the multiplication laws for the additional
parts $o'_I$ algebra  $\mathcal{A}'(Y(k), AdS_d)$  and for
converted operators $O_I$ algebra $\mathcal{A}_c(Y(k), AdS_d)$.

As the new result we write down  the explicit form for the cubic
commutator  algebras $\mathcal{A}'$, $\mathcal{A}_c$
\begin{eqnarray}
 [\,o_I',o_J'] & =& f_{IJ}^{(1)K_1}{o}'_{K_1}-f_{IJ}^{(2)K_{2}K_1}{o}'_{K_1}{o}'_{K_2}
+ f_{IJ}^{(3)K_{3}K_2 K_1}{o}'_{K_1}{o}'_{K_2}{o}'_{K_3},\nonumber \\
\label{conv-alg3}
   [\,{O}_I,{O}_J] & = &
  \Bigl(f_{IJ}^{(1)K} + \sum_{l=2}^{n}
  F^{(l){}K}_{IJ}({o}',{O})\Bigr){O}_{K},\\
 F^{(3){}K}_{IJ}  &=&f_{IJ}^{(3)K_1K_{2}K}{O}_{K_1}{O}_{K_2}-
\bigl(f_{IJ}^{(3)K_1K_2K} +
f_{IJ}^{(3)K_2K_{1}K}\bigr){o'}_{K_1}{O}_{K_2}\nonumber\\
{}&& + \bigl(f_{IJ}^{(3)K_2K_1K} +
 f_{IJ}^{(3)K_2KK_{1}} +
f_{IJ}^{(3)KK_2K_{1}}\bigr){o'}_{K_1}{o'}_{K_2},
\end{eqnarray}
if the commutator relations for the initial algebra $\mathcal{A}$
given by the Eqs.(\ref{polynom}) for  $n=3$.


\section{Auxiliary HS symmetry algebra $\mathcal{A}'(Y(2),
AdS_d)$}\label{HSauxY2b}

The procedure of additive conversion for non-linear HS symmetry
algebra $\mathcal{A}(Y(k), AdS_d)$ of the operators  $o_I$ implies
finding, first,  the explicit form of the algebra
$\mathcal{A}'(Y(k), AdS_d)$ of the additional parts  $o'_I$,
second, the representation of $\mathcal{A}'(Y(k), AdS_d)$ in terms
of some appropriate Heisenberg algebra elements acting in a new
Fock space $\mathcal{H}'$.  Structure of the non-linear
commutators of the initial algebra leads to necessity to  convert
all the operators $o_I$ to construct unconstrained LF for given HS
field $\Phi_{(\mu^1)_{s_1},(\mu^2)_{s_2},...,(\mu^k)_{s_k}}$.

Considering here the case of $k=2$ family of indices only in the initial
 HS field $\Phi_{(\mu^1)_{s_1},(\mu^2)_{s_2}}$  (the general case of algebra $\mathcal{A}'(Y(k), AdS_d)$ is discussed in \cite{BurdikReshetnyak})
we see the former step is based on a determination of a multiplication table
$\mathcal{A}'(Y(2), AdS_d)$ of operators $o'_I$ following to the
form of the algebra $\mathcal{A}(Y(k), AdS_d)$ for $k=2$ given by the
table~\ref{table} and Eqs.(\ref{auxalg2}).

 As the result, the
searched composition law for $\mathcal{A}'(Y(2), AdS_d)$ is the
same as for the algebra $\mathcal{A}(Y(2), AdS_d)$ in its  linear
Lie part, i.e. for
 $sp(4)$ subalgebra of elements $(l^{\prime ij}, l^{\prime ij+},
t^{\prime 12}, t^{\prime+}_{12},g_0^{\prime i})$\footnote{To turn from
general algebra $\mathcal{A}(Y(k), AdS_d)$ to $\mathcal{A}(Y(2), AdS_d)$,
we   put
$\theta^{ij}= \delta^{i2}\delta^{j1}$ and therefore only surviving
operators among mixed-symmetry ones are $t^{ij} = t^{12}$, $t^+_{ij} =  t^{ +}_{12}$.
},  and is differed in
the non-linear part of the Table~\ref{table}, determined by the
isometry group elements $l'_i, l^{\prime +}_j, l'_0$. The
corresponding non-linear submatrix of the multiplication matrix
for $\mathcal{A}'(Y(2), AdS_d)$ has the form given by the
Table~\ref{table'},
\begin{table}[t]{\caption{The non-linear part of  algebra $\mathcal{A}'(Y(2),
AdS_d)$.} \label{table'}
\begin{center}
\begin{tabular}{||c||c|c|c||}\hline\hline
$[\,\downarrow\,,\to\}$&
 $l'_0$ &
$l^{\prime i}$ & $l^{\prime i{}+}$  \\
\hline\hline $l'_0$
    & $0$
   &
    $ r{\mathcal{K}}^{\prime bi+}_1$ & $-r{\mathcal{K}}^{\prime i}_1$\\
\hline $l^{\prime j}$
   &   $ -r{\mathcal{K}}^{\prime j+}_1$
   & $-{W}^{\prime ji}_b$  & ${X}^{\prime ji}_b$ \\
\hline $l^{\prime j+}$ &
   $r{\mathcal{K}}^{\prime j}_1$  &
   $-{X}^{\prime ij}_b$
   & $ {W}^{\prime ji+}_b$  \\
   \hline\hline
\end{tabular}\end{center}
}
\end{table}
Here the functions ${\mathcal{K}}^{\prime i+}_1$,
${\mathcal{K}}^{\prime i}_1$, ${W}^{\prime ji}_b$, ${W}^{\prime
ji+}_b$ $(X^{\prime ij}_b-l'_0)$ have the same definition as the
ones (\ref{lilj})--(\ref{lilj+b}) for initial operators $o_I$ but
with opposite sign for $(X^{\prime ij}_b-l'_0)$ and for $k=2$:
\begin{eqnarray}
{} {W}^{\prime ij}_b & = & 2r\epsilon^{ij} \left[(g_0^{\prime
2}-g_0^{\prime 1})l^{\prime 12} - t^{\prime 12}l^{\prime 11}+
t^{\prime 12+}l^{\prime 22} \right],\
 \label{lilj'2}\\
{\mathcal{K}}^{\prime j}_1& = &\Bigl(4\sum_{i}l^{\prime ji+}
l^{\prime i} +l^{\prime j+}(2g_0^{\prime j}-1) -2l^{\prime
2+}t^{\prime +12}\delta^{j1} -2l^{\prime 1+}t^{\prime
12}\delta^{j2}\Bigr)  \label{l'0li+'2} ,
\\
 {} {X}^{\prime ij}_b
  & =
 &\bigl\{l'_{0}- r\bigl( K^{\prime 0i}_0  +\mathcal{K}^{ \prime 12}_0\bigr)\bigr\}\delta^{ij}
      +
r\bigl\{ 4\sum\nolimits_{l} l^{\prime 1l+}l^{\prime l2}
 +
\textstyle(g_0^{\prime
 2}+g_0^{\prime 1}
  -2)t'_{ 12}\bigr\}\delta^{i2}\delta^{j1}\nonumber\\
&& + r\bigl\{ 4\sum\nolimits_{l} l^{\prime l2+}l^{\prime 1l}
+t^{\prime +}_{12}(g_0^{\prime 2}+g_0^{\prime 1}-2
)\bigr\}\delta^{i1}\delta^{j2},  \label{lilj+b'2}
\end{eqnarray}
with totally
    antisymmetric $sp(2)$-invariant tensor
$\epsilon^{ij}$, $\epsilon^{12}=1$ and  operator $\mathcal{K}^{
12}_0 = \bigl(t^{ +}_{12}t^{12}- 4l^{ +}_{12}
      l^{ {12}} -
    g_0^{ 2} \bigr)$  derived from Casimir operator for $sp(4)$ algebra
    in the Eqs.(\ref{Casimirsb}) for $k=2$.
In turn,
the Lie part of the
Tables~\ref{table}  for $k=2$ is the same as one for
bosonic Lie subalgebra in \cite{flatfermmix} for the following
expressions of only non-vanishing operators  $B^{\prime
12}{}_{12}$, $A^{\prime 12{}12}$, $F^{\prime 12,j}$, $F^{\prime
12,j+}$,  $L^{\prime i_1j_1,i_2j_2}$
\begin{eqnarray}
&& B^{\prime 12}{}_{12} = (g_0^{\prime 1} - g_0^{\prime 2}),\quad
A^{\prime 12,12} =0, \quad  F^{\prime 12,j} \equiv  F^{\prime j}=
t^{\prime }_{12}(
\delta^{j2}-\delta^{j1}),  \label{B1212}\\
&& L^{\prime i_1j_1,i_2j_2} =
\textstyle\frac{1}{4}\Bigl\{\delta^{i_2i_1}
\delta^{j_2j_1}\Bigl[2g_0^{\prime i_2}\delta^{i_2j_2} +
g_0^{\prime i_2} + g_0^{\prime j_2}\Bigr]  -
\delta^{j_2\{i_1}\Bigl[t^{\prime }_{12}
\delta^{j_1\}1}\delta^{i_22}
 +t^{\prime +}_{12}\delta^{j_1\}2}\delta^{i_21}\Bigr] \nonumber \\
&& \hspace{2em} -
\delta^{i_2\{i_1}\Bigl[t^{\prime}_{12}\delta^{j_22}\delta^{j_1\}1}
+t^{\prime +}_{12}\delta^{j_1\}2}\delta^{j_21}\Bigr] \Bigr\}
 \,,\label{Lklij2}
\end{eqnarray}
 being identical to the same operators $o_I$  from the initial
algebra $\mathcal{A}(Y(2), AdS_d)$.

Now, we may to sketch the points to find oscillator representation
for the elements $o'_I$ of auxiliary HS symmetry algebra.
\subsection{Verma module for the quadratic algebra $\mathcal{A}'(Y(2k),AdS_d)$}
\label{VMboconic}
 Here, we following to assumption that
generalization of Poincare--Birkhoff--Witt theorem for the second
order algebra $\mathcal{A}'(Y(2),AdS_d)$ is true (see on  PBW
theorem generalization for quadratic algebra
\cite{PBWtheorem_qa}), we start to construct Verma module, based
on Cartan-like decomposition enlarged from one for $sp(4)$
($i\leq j$)
    \begin{eqnarray}\label{Cartandecomp2}
 \hspace{-1em}&&  \hspace{-1em} \mathcal{A}'(Y(2),AdS_d) =  \{l^{\prime +}_{ij},
t^{\prime+}_{12}, l^{\prime +}_i\} \oplus \{g_0^{\prime i}, l_0'\}
\oplus \{l^{\prime }_{ij}, t'_{12}, l^{\prime }_i\} \equiv
\mathcal{E}^-_2\oplus H_2 \oplus\mathcal{E}^+_2.\footnotemark
\end{eqnarray}
\footnotetext{we may consider $sp(4)$ and generally $sp(2k)$ in
Cartan-Weyl basis for
unified description, however without loss of generality the basis
elements and structure constants of the algebra under
consideration will be chosen as in the table~\ref{table}}Note,
that in contrast to the case of Lie algebra the element $l'_0$
does not diagonalize the elements of upper $\mathcal{E}^-_2$
(lower $\mathcal{E}^+_2$) triangular subalgebra due to quadratic
relations (\ref{l'0li+'2}) (as well as it was for totally-symmetric
HS fields on AdS space \cite{BurdikNavratilPasnev, BKL}) and in addition   the negative
root vectors $l^{\prime +}_i, l^{\prime +}_j$ do not commute.

Because the Verma module over a semi-simple finite-dimensional Lie
algebra $g$  (an induced module
$\mathcal{U}(g)\bigotimes_{\mathcal{U}(b)} |0\rangle_V $ with a
vacuum vector $|0\rangle_V$\footnote{here the signs
$\mathcal{U}(g)$, $\mathcal{U}(b)$, $\mathcal{U}(g^{-})$ denote
the universal enveloping algebras respectively for $g$, for its
Borel subalgebra and for lower triangular subalgebra $g^-$ like
$\mathcal{E}^-_2$ in (\ref{Cartandecomp2})}) is isomorphic due to
PBW theorem as a vector space to a polynomial algebra
$\mathcal{U}(g^{-})\bigotimes_{C} |0\rangle_V $, it is clear that
$g$ can be realized by first-order inhomogeneous differential
operators acting on these polynomials.

We consider generalization of \emph{Verma module notion for the
case of quadratic algebras} $g(r)$ which present the
$g(r)$-deformation of Lie algebra $g$ in such form that  $g(r=0) =
g$. Thus, we consider Verma module for such kinds of non-linear
algebras, supposing that PBW theorem is valid for $g(r)$ as well
that will be proved later by the explicit construction of the
Verma module.

Doing so, we consider the quadratic algebra
$\mathcal{A}'(Y(2),AdS_d)$, as $r$-deformation of Lie algebra
$\mathcal{A}'(Y(2),R^{1,d-1})$ with use of (\ref{identalg}) for $k=2$ (see
for details \cite{BuchbinderReshetnyak}),
\begin{equation}\label{identalgaux} \mathcal{A}'(Y(2),AdS_d) = \left(T^{\prime 2} \oplus
T^{\prime 2*}\oplus l'_0\right)(r) + \hspace{-1em} \supset
  sp(4),\quad T^{\prime 2}
= \bigl\{ l'_i\bigr\},\  T^{\prime 2*} = \bigl\{ l^{\prime
+}_i\bigr\}.
 \end{equation}
Corresponding basis vector of  Verma module
$|\vec{N}(2)\rangle_V$ has, therefore the form according to
(\ref{Cartandecomp2})
\begin{eqnarray}   |\vec{N}(2)\rangle_V = \left| {\vec{n}}_{ij},{n}_{1}, {p}_{12},n_2\rangle_V
\right. \equiv \prod_{i\leq j}^2\textstyle\bigl(l^{\prime
+}_{ij}\bigr){}^{n_{ij}}\left(\frac{l^{\prime
+}_1}{m_1}\right){}^{ n_1}\bigl(t^{\prime
+}_{12}\bigr){}^{p_{12}}\left(\frac{l^{\prime
+}_2}{m_2}\right){}^{ n_2} |0\rangle_V, \
\mathcal{E}^+_2|0\rangle_V=0, \label{N2}
\end{eqnarray}
 with (vacuum) highest weight vector
$|0\rangle_V$, non-negative integers  $n_{ij}, n_l, p_{lm}$, and
arbitrary constants $m_l$ with dimension of mass.

Here, in contrast to the case of Lie algebra \cite{flatfermmix} and totally-symmetric
HS fields on AdS space \cite{BurdikNavratilPasnev}, \cite{adsfermBKR},
the negative root vectors $l^{\prime +}_1, t^{\prime+}_{12}, l^{\prime
+}_2$ do not commute, making the
 vector $|\vec{N}(2)\rangle_V
$ by not proper one for the operators $t^{\prime+}_{12}, l^{\prime
 +}_2$ presenting therefore  the essential peculiarities to construct
Verma module for  $\mathcal{A}'(Y(2),AdS_d)$.

First of all, by the definition the highest weight vector
$|0\rangle_V$ is proper for the vectors from the Cartan-like
subalgebra $H_2$,
\begin{equation}\label{nullV}
   (g_0^{\prime i}, l_0')|0\rangle_{V} =
 (h^i, m_0^2) |0\rangle_{V} ,
\end{equation}
with some real numbers $h^1, h^2, m_0$ whose values will be
determined later in the end of LF construction in.
 order to Lagrangian equations of motion reproduce the initial
 AdS group irreps conditions (\ref{Eq-0b})--(\ref{Eq-3b}).

Then, we determine the action of  the negative root vectors from
the subspace $\mathcal{E}^-_2$ on the basis vector
$|\vec{N}(2)\rangle_V$ which now do not present explicitly the
action of raising operators  and reads,
\begin{eqnarray}
l^{\prime+}_{lm}\left|\vec{N}(2)\rangle_V \right. & = &
\left|\vec{n}_{ij}+ \delta_{ij,lm}, \vec{n}_s
 \rangle_V \right. \,,
 \label{l'+ijb} \\
\label{l'+lb}
 l^{\prime  +}_l\left|\vec{N}(2)\rangle_V
 \right. & = & \delta^{l2}
 \prod_{i\leq j}^2\textstyle\bigl(l^{\prime
+}_{ij}\bigr){}^{n_{ij}} l^{\prime +}_2\left|
\vec{0}_{ij},\vec{n}_s\rangle_V\right.  +\delta^{l1}
m_1\left|\vec{n}_{ij}, n_1 +1,
 p_{12}, n_2\rangle_V\right. \,,\ l=1,2,
 \\
 t^{\prime+}_{12}  \left|\vec{N}(2)\rangle_V
 \right. & = & \prod_{i\leq j}^2\textstyle\bigl(l^{\prime
+}_{ij}\bigr){}^{n_{ij}}t^{\prime
+}_{12}|\vec{{0}}_{ij},\vec{n}_s\rangle_V -
 2n_{11}\left| {n}_{11}-1, n_{12}+1, n_{22}, \vec{n}_s\rangle_V
 \right. \nonumber\\
  &&  - n_{12}\left| {n}_{11}, {n}_{12}-1, {n}_{22}+1,
\vec{n}_s\rangle_V
  \right. .
 \label{t'+b}
\end{eqnarray}
 Here we have used the
notations, first, $\vec{n}_s \equiv (n_1, p_{12}, n_2)$,
$\vec{0}_{ij} \equiv (0,0,0)$ in accordance with definition of
$|\vec{N}(2)\rangle_V$, second,
$\delta_{ij,lm}=\delta_{il}\delta_{jm}$, for $i\leq j, l\leq m$,
so that the vector $\left|\vec{N} +
\delta_{ij,lm}\rangle_V\right.$ in the Eq.(\ref{l'+ijb}) means
subject to definition (\ref{N2}) increasing of only the
coordinate $n_{ij}$ in the vector $|\vec{N}(2)\rangle_V$, for $i=l,
j=m$, on unit with unchanged values of the rest ones.

In turn, the action of Cartan-like generators on the vector
$|\vec{N}(2)\rangle_V
 $ are
given by the relations,
\begin{eqnarray}
  %
 g_0^{\prime l} \left|\vec{N}(2)\rangle_V
 \right. & = & \Bigl( 2n_{ll}+n_{12} + n_l + (-1)^l p_{12} + h^l\Bigr)
 \left|\vec{N}(2)\rangle_V
 \right.,\ l=1,2, \label{g'0ib}\\
 \label{l'0auxb}
 l'_0\left|\vec{N}(2)\rangle_V
 \right. & =& \prod\nolimits_{i\leq j}^2\textstyle\bigl(l^{\prime
+}_{ij}\bigr){}^{n_{ij}}l'_0\left| \vec{0}_{ij},
\vec{n}_s\rangle_V\right..
\end{eqnarray}
To derive the Eqs. (\ref{l'+ijb})--(\ref{l'0auxb}) we have used
the formula for the product of  operators $A$, $B$,
\begin{eqnarray}
\label{product}     AB^n & = &\sum^{n}_{k=0} C^n_k
B^{n-k}\mathrm{ad}^k_B{}A\,,   \ \mathrm{ad}^k_B{}A=
[[...[A,\stackrel{ k{\,} {\rm times}}{
\overbrace{B\},...\},B}\}}, \verb" for " n\geq 0, C^n_k=\frac{n!}{k!(n-k)!}.
\end{eqnarray}
At last,  the action of the positive root vectors from the
subspace $\mathcal{E}^+_2$   on the vector $|\vec{N}(2)\rangle_V$
being based on  the rule (\ref{product}), reads as follows,
\begin{eqnarray}
\label{t'auxb}
  t^{\prime}_{12}
\left|\vec{N}(2)\rangle_V
 \right. &=&  \hspace{-0.3em}  \prod_{i\leq j}^2\textstyle\bigl(l^{\prime
+}_{ij}\bigr){}^{n_{ij}}\hspace{-0.3em}\left(\frac{l^{\prime
+}_1}{m_1}\right)\hspace{-0.2em}{}^{ n_1}\bigl(t^{\prime
+}_{12}\bigr){}^{p_{12}} t'_{12}\left| \vec{0}_{ij},
0,0,n_2\rangle_V \right. -\sum\limits_l
ln_{l2}\hspace{-0.2em}\left|{\vec{n}}_{ij}-\delta_{ij,l2}+
\delta_{ij,1l}, \vec{n}_s\rangle_V
 \right. \nonumber \\
  &&+
p_{12}(h^1-h^2-n_2-p_{12}+1)\left|\vec{n}_{ij}, n_1, p_{12}-1,
{n}_2\rangle_V
 \right.
\,,\\
\label{l'1auxb}
 l^{\prime 1}\left|\vec{N}(2)\rangle_V
 \right. & =& -m_1 n_{11}\bigl|{\vec{n}}_{ij}-\delta_{ij,11},
  n_1+1, p_{12}, {n}_2\rangle_V \nonumber\\
{}&{}&+\Bigl\{ \prod_{i\leq j}^2\bigl(l^{\prime
+}_{ij}\bigr){}^{n_{ij}}l^{\prime 1}- \frac{n_{12}}{2}\prod_{i\leq
j}^2\bigl(l^{\prime +}_{ij}\bigr){}^{n_{ij}-\delta_{ij,12}}
l^{\prime  +}_2 \Bigr\}\left|\vec{{0}}_{ij},
\vec{n}_s\rangle_V\right.,
 \\
\label{l'2auxb}
 l^{\prime 2}\left|\vec{N}(2)\rangle_V
 \right. & =& -m_1 \frac{n_{12}}{2}\bigl| {\vec{n}}_{ij}-\delta_{ij,12},
   n_1+1, p_{12}, {n}_2\rangle_V \nonumber \\
{}&{}& + \Bigl\{ \prod_{i\leq j}^2\bigl(l^{\prime
+}_{ij}\bigr){}^{n_{ij}}l^{\prime 2} -{n_{22}} \prod_{i\leq
j}^2\bigl(l^{\prime
+}_{ij}\bigr){}^{n_{ij}-\delta_{ij,22}}l^{\prime +}_2
\Bigr\}\left|\vec{{0}}_{ij},
\vec{n}_s\rangle_V\right.,\nonumber\\
  {}\label{l'11auxb}
  l^{\prime 11}
\hspace{-0.2em}\left|\vec{N}(2)\rangle_V
 \right. \hspace{-0.2em}&\hspace{-0.2em}=\hspace{-0.2em}&
 \hspace{-0.2em}
   {n}_{11}({n}_{11}+{n}_{12}+n_1- p_{12} -1 + h^1)
   \left|{\vec{n}}_{ij} -\delta_{ij,11}, \vec{n}_s\rangle_V
 \right.\nonumber \\ &&+ \frac{{n}_{12}({n}_{12}-1)}{4}
 \left| {\vec{n}}_{ij} -2\delta_{ij,12}+\delta_{ij,22}
, \vec{n}_s\rangle_V
 \right.\nonumber
  \\
 && \hspace{-0.2em}
 +\Bigl\{\prod_{i\leq j}^2\bigl(l^{\prime
+}_{ij}\bigr){}^{n_{ij}}l^{\prime 11} -
\frac{{n}_{12}}{2}\prod_{i\leq j}^2\bigl(l^{\prime
+}_{ij}\bigr){}^{n_{ij}-\delta_{ij,12}}t^{\prime +}_{12}
\Bigr\}\left|\vec{{0}}_{ij},
\vec{n}_s\rangle_V\right.,\\
     l^{\prime 12}
\left|\vec{N}(2)\rangle_V
 \right. &=&  \frac{{n}_{12}}{4}
\Bigl({n}_{12}+
 \displaystyle\sum_{l}\hspace{-0.1em}(2{n}_{ll}+n_l+h^l)-1
\Bigr)\left|
 {\vec{n}}_{ij} - \delta_{ij,12},  \vec{n}_s\rangle_V
 \right.\nonumber
 \\
&& + \frac{1}{2} p_{12} n_{11}(h^2-h^1 + n_2+ p_{12} - 1)\left|
 {\vec{n}}_{ij}-\delta_{ij,11}, n_1, p_{12}-1, {n}_2\rangle_V \right.
 \nonumber \\
 &&  + n_{11}n_{22}\left|
 {\vec{n}}_{ij}-\delta_{ij,11} -\delta_{ij,22}+ \delta_{ij,12}, \vec{n}_s\rangle_V
 \right.
\nonumber \\
 && +\Bigl\{\prod_{i\leq j}^2\bigl(l^{\prime
+}_{ij}\bigr){}^{n_{ij}}l^{\prime 12}-
\frac{n_{22}}{2}\displaystyle\prod_{i\leq j}^2\bigl(l^{\prime
+}_{ij}\bigr){}^{n_{ij}-\delta_{ij,22}}t^{\prime
+}_{12}\Bigr\}\left|\vec{{0}}_{ij}, \vec{n}_s\rangle_V\right. \nonumber \\
&& -\frac{n_{11}}{2}\displaystyle\prod_{i\leq j}^2\bigl(l^{\prime
+}_{ij}\bigr){}^{n_{ij}-\delta_{ij,11}}\textstyle\left(\frac{l^{\prime
+}_1}{m_1}\right){}^{ n_1}\bigl(t^{\prime
+}_{12}\bigr){}^{p_{12}}t^{\prime
}_{12}\left|\vec{{0}}_{ij},0,0, {n}_2\rangle_V\right.\label{l'12auxb}\\
  l^{\prime 22}
\left|\vec{N}(2)\rangle_V
 \right. &=&  n_{22}(n_{12}+ n_{22} + p_{12}  + n_2  -1 + h^2)
 \left|{\vec{n}}_{ij}-\delta_{ij,22}, \vec{n}_s\rangle_V
 \right. \nonumber\\
  && +  \frac{n_{12}p_{12}}{2}(p_{12}-1 + h^2 - h^1+n_2)\left|
{\vec{n}}_{ij}-\delta_{ij,12},
 n_1, p_{12}-1, {n}_2\rangle_V
 \right.
\nonumber \\
 && + \frac{n_{12}(n_{12}-1)}{4}\left|{\vec{n}}_{ij}+\delta_{ij,11}
-2\delta_{ij,12},
  \vec{n}_s\rangle_V
 \right.
\nonumber \\
 && +\prod_{i\leq j}^2\bigl(l^{\prime
+}_{ij}\bigr){}^{n_{ij}}\hspace{-0.3em}
\textstyle\left(\frac{l^{\prime
+}_1}{m_1}\hspace{-0.3em}\right){}^{ n_1}l^{\prime
22}\left|\vec{0}_{ij}, 0,p_{12}, {n}_2\rangle_V\right.
 \nonumber\\
&&- \displaystyle\frac{n_{12}}{2}\prod_{i\leq j}^2\bigl(l^{\prime
+}_{ij}\bigr){}^{n_{ij}-\delta_{ij,12}}
\textstyle\left(\frac{l^{\prime +}_1}{m_1}\right){}^{
n_1}\bigl(t^{\prime +}_{12}\bigr){}^{p_{12}}t^{\prime}_{12}
\left|\vec{0}_{ij}, 0,0, {n}_2\rangle_V\right. \,.\label{l'22auxb}
\end{eqnarray}
It is easy to see that to complete the  calculations  in Eqs.
(\ref{l'+lb}), (\ref{t'+b}), (\ref{l'1auxb})--(\ref{l'22auxb})
 we need to find the  action of positive root vectors
 $ l^{\prime m}, l^{\prime 1m}$, $m=1,2$, Cartan-like
vector $l^{\prime }_0$, negative root vectors $l^{\prime +}_2,
t^{\prime +}_{12}$ on the  vector $|\vec{{0}}_{ij},
\vec{n}_s\rangle_V.$
 and the rest operators $t'_{12},  l^{\prime 22}$
on an arbitrary vector $|\vec{0}_{lm},0,0,{n}_2\rangle_V$ in terms
of linear combinations of  definite vectors. We  solve this
rather non-trivial technical problem   explicitly
 with introducing auxiliary quantities which we
call \emph{primary   block-operator} $\widehat{t}'_{12}$ given  in
(\ref{t'ntot}) and \emph{derived block-operators}
$\widehat{t}^{\prime +}_{12}, \widehat{l}^{\prime +}_{2},
\widehat{l}^{\prime }_{0} , \widehat{l}^{\prime }_{m},
 \widehat{l}^{\prime }_{m2}, m=1,2$ below whose concrete expressions will be
shown with details of Verma module construction for the algebra under
consideration in \cite{BurdikReshetnyak}.
Thus, we  may to formulate the result in the form~of

\vspace{1ex} \noindent \textbf{\emph{Theorem 1.}} The Verma module
for the  non-linear second order algebra $\mathcal{A}'(Y(2),
AdS_d)$ exists, is determined by the relations (\ref{l'+ijb}),
(\ref{g'0ib}), 
(\ref{t12fin})--(\ref{l22totnV}), expressed with help of primary
$\widehat{t}'_{12}$  and derived block-operators
$\widehat{t}^{\prime +}_{12}, \widehat{l}^{\prime +}_{2},
\widehat{l}^{\prime }_{0} , \widehat{l}^{\prime }_{m},
 \widehat{l}^{\prime }_{m2}, m=1,2$
 and has the final form,
\begin{eqnarray}
\label{t12fin} t'_{12}\left|\vec{N}(2)\rangle_V\right. & =&
 p_{12}(h^1-h^2-n_2-p_{12}+1)\left|\vec{n}_{ij}, n_1, p_{12}-1, {n}_2\rangle_V
 \right. \nonumber \\
  && -\sum_l ln_{l2}\hspace{-0.2em}\left|{\vec{n}}_{ij} -
\delta_{ij,l2}+ \delta_{ij,1l}, \vec{n}_s\rangle_V
 \right.
+ \hat{t}'_{12} \left|\vec{N}(2)\rangle_V\right.,\\
{}
 {{t}^{\prime +}_{12}\left|\vec{N}(2)\rangle_V\right.} & =& -
 \sum_l (3-l) n_{1l}\left| {\vec{n}}_{ij}-\delta_{ij,1l} +\delta_{ij,l2},
 \vec{n}_s\rangle_V
 \right. +\hat{t}^{\prime +}_{12}\left|\vec{N}(2)\rangle_V\right.,
 \label{t+fin2}\\
 {{l}^{\prime
+}_l\left|\vec{N}(2)\rangle_V\right.} & = &\delta^{l1}
m_1\left|\vec{N}(2) + \delta_{s,1}\rangle_V\right. +
\delta^{l2}\hat{l}^{\prime +}_{2}\left|\vec{N}(2)\rangle_V\right.,
\label{ll+fin2}\\
  {l^{\prime  }_0\left|\vec{N}(2)\rangle_V\right.}
 &=& {{l}^{\prime  }_0\left|\vec{0}_{ij},\vec{n}_s\rangle_V\right.}\vert_{[\vec{0}_{ij}\to
 \vec{n}_{ij}]},   \label{l0finnV}\\
 {l^{\prime  }_1\left|\vec{N}(2)\rangle_V\right.}
& =& {l^{\prime
}_1\left|\vec{0}_{ij},\vec{n}_s\rangle_V\right.}\vert_{[
\vec{0}_{ij}\to
 \vec{n}_{ij}]}  -m_1 n_{11}\left|\vec{n}_{ij}-\delta_{ij,11},
  \vec{n}_s+\delta_{s,1}\rangle_V\right. \nonumber \\
{}&{}&- \frac{n_{12}}{2}\widehat{l}^{\prime
+}_2\left|\vec{n}_{ij}-\delta_{ij,12},\vec{n}_s\rangle_V\right. ,
\label{l1finnV}
\\
 {l^{\prime
}_2\left|\vec{N}(2)\rangle_V\right.} & = & {l^{\prime
}_2\left|\vec{0}_{ij},\vec{n}_s\rangle_V\right.}\vert_{
[\vec{0}_{ij}\to
 \vec{n}_{ij}]} -m_1 \frac{n_{12}}{2}
 \left| {\vec{n}}_{ij}-\delta_{ij,12},
   \vec{n}_s +\delta_{s,1}\rangle_V\right. \\
{}&{}& - {n_{22}} {\widehat{l}^{\prime +
}_2\left|\vec{n}_{ij}-\delta_{ij,22},\vec{n}_s\rangle_V\right.},
\label{l2finnV}
\\
 {l^{\prime
}_{11}\left|\vec{N}(2)\rangle_V\right.} &=&
  {n}_{11}({n}_{11}+{n}_{12}+n_1- p_{12} -1 + h^1)
   \left|\vec{n}_{ij} -\delta_{ij,11}, \vec{n}_s\rangle_V
 \right. \nonumber\\
 && + \frac{{n}_{12}({n}_{12}-1)}{4}
 \left| \vec{n}_{ij} -2\delta_{ij,12}+\delta_{ij,22},
\vec{n}_s\rangle_V \right.+l^{\prime
}_{11}\left|\vec{0}_{ij},\vec{n}_s\rangle_V\right.\vert_{[\vec{0}_{ij}
\to \vec{n}_{ij}]}
 \nonumber \\
 &&
 - \frac{{n}_{12}}{2}\widehat{t}^{\prime
+}_{12}\left|\vec{n}_{ij}-\delta_{ij,12},\vec{n}_s\rangle_V
 \right.,
 \label{l11nVfin}
\\
 {l^{\prime }_{12}\left|\vec{N}(2)\rangle_V\right.}& =
&\frac{{n}_{12}}{4} \Bigl({n}_{12}+
 \displaystyle\sum_{l}\hspace{-0.1em}(2{n}_{ll}+n_l+h^l)-1
\Bigr)\left|
 {\vec{n}}_{ij} - \delta_{ij,12},  \vec{n}_s\rangle_V
 \right.\nonumber
 \\
&& + \frac{1}{2} p_{12} n_{11}(h^2-h^1 + n_2+ p_{12} - 1)\left|
 {\vec{n}}_{ij}-\delta_{ij,11}, n_1, p_{12}-1, {n}_2\rangle_V \right.
 \nonumber \\
 &&  + n_{11}n_{22}\left|
 {\vec{n}}_{ij}-\delta_{ij,11} -\delta_{ij,22}+ \delta_{ij,12}, \vec{n}_s\rangle_V
 \right. + l^{\prime
}_{12}\left|\vec{0}_{ij},\vec{n}_s\rangle_V\right.\vert_{[\vec{0}_{ij}
\to
\vec{n}_{ij}]} \nonumber\\
&& -\frac{n_{22}}{2}\widehat{t}^{\prime
+}_{12}\left|\vec{n}_{ij}-\delta_{ij,22},\vec{n}_s\rangle_V\right.-\frac{n_{11}}{2}\widehat{t}^{\prime
}_{12}\left|\vec{n}_{ij}-\delta_{ij,11},\vec{n}_s\rangle_V\right.,
 \label{l12nVfin}
\\
{l^{\prime }_{22}\left|\vec{N}(2)\rangle_V\right.} &= &
n_{22}(n_{12} + p_{12} + n_2 + n_{22} -1 + h^2)
 \left|\vec{n}_{ij} -\delta_{ij,22}, \vec{n}_s\rangle_V
 \right. \nonumber\\
  && +  \frac{n_{12}p_{12}}{2}(p_{12}-1 + h^2 - h^1+n_2)
\left| \vec{n}_{ij}-\delta_{ij,12},
 \vec{n}_s -\delta_{s,12}\rangle_V
 \right.
\nonumber \\
 && +  \frac{n_{12}(n_{12}-1)}{4}\left|\vec{n}_{ij}+\delta_{ij,11}
-2\delta_{ij,12}, \vec{n}_s\rangle_V\right.+\widehat{l}^{\prime
}_{22}\left|\vec{N}(2)\rangle_V\right. \nonumber\\
&&
-\frac{n_{12}}{2}\widehat{t}'_{12}\left|{\vec{n}}_{ij}-\delta_{ij,12},
   \vec{n}_s\rangle_V\right. \label{l22totnV}
.
\end{eqnarray}
In deriving these relations the rule was used
\begin{equation}\label{convnsN2}
\prod_{i\leq j}^2\bigl(l^{\prime +}_{ij}\bigr){}^{n_{ij}} (l'_0,
l'_n, l'_{lm})\left|\vec{0}_{ij},\vec{n}_s\rangle_V\right. \equiv
(l'_0, l'_n, l'_{lm})\left|\vec{0}_{ij},\vec{ n}_s
\rangle_V\right.\vert_{[\vec{0}_{ij}\to \vec{n}_{ij}]},\quad
l,m,n=1,2, \ l\leq m,
\end{equation}
when the multipliers $\bigl(l^{\prime +}_{ij}\bigr){}^{n_{ij}}$
act as only raising operators on  vectors $(l'_0, l'_k,
l'_{km})\left|\vec{0}_{ij},\vec{n}_s\rangle_V\right.$.
The \emph{primary block operator} $\hat{t}'_{12}$
(corresponding to non-Lie part of $t'_{12}$) is
determined as follows  \begin{eqnarray}
\label{t'ntot}
 \hat{t}'_{12}\left|\vec{N}(2)\rangle_V\right. &=&
   \sum\limits^{\left[(n_2-1)/2\right]}_{k=0}\Biggl\{\sum\limits^{
  \left[n_2/2\right]}_{{}^{1}m=0}
  \sum\limits^{\left[n_2/2-({}^{1}m + 1)\right]}_{{}^1l=0}\ldots\nonumber \\
\hspace{-1em}&&\hspace{-1em}
  \ldots\sum\limits^{\left[n_2/2-\sum\limits^{k-1}_{i=1}({}^{i}m + {}^il)
  -(k-1)\right]}_{{}^{k}m=0}
  \sum\limits^{\left[n_2/2-\sum\limits^{k-1}_{i=1}({}^{i}m + {}^il)-{}^{k}m
  -k \right]}_{{}^kl=0} (-1)^k\left(\frac{-2r}{m_2^2}\right)^{\sum_{i=1}^k({}^im+{}^il)+k}
\nonumber\\
\hspace{-1em}&&\hspace{-1em} \times C^{n_2}_{2{}^1m+1}C^{n_2-2{}^1m-1}_{2{}^1l+1}...
C^{n_2-2(\sum_{i=1}^{k-1}({}^im+{}^il)+k-1)}_{2{}^km+1}C^{n_2-2(
\sum_{i=1}^{k-1}(
{}^im+{}^il) +k-1)-2{}^km-1}_{2{}^kl+1}\nonumber\\
&&\ \ \times
\Big|\hat{A}_{\vec{n}_{ij}+[\sum\limits^k_{i=1}({}^{i}m + {}^il)
  +k]\delta_{i2}\delta_{j2},{n}_1,n, n_2-2[\sum\limits^k_{i=1}({}^{i}m + {}^il)
  +k]}\Big\rangle_V\Biggr\},
\end{eqnarray}
with the vector $\left|\hat{A}_{\vec{N}(2)}\rangle_V\right.$ given as,
\begin{eqnarray}
 \left|\hat{A}_{\vec{N}(2)}\rangle_V\right. & =&
 n_2p_{12} \left|\vec{N}(2) -\delta_{s,12}\rangle_V\right.\label{Antot}
\nonumber\\
&\hspace{-1em}&\hspace{-1em}  -\frac{m_1}{m_2}
\sum\limits^{\left[n_2/2\right]}_{{}^1m=0}\left(\frac{-2r}{
    m_2^2}\right)^{{}^1m}C^{n_2}_{2{}^1m +1}\left|\vec{N}(2)+{}^1m
    \delta_{ij,22}+\delta_{s,1}-(2{}^1m+1)\delta_{s,2}
    \rangle_V\right.  \nonumber\\
        &\hspace{-1em}&\hspace{-1em} -\sum\limits^{\left[n_2/2\right]}_{{}^1m=1}\left(\frac{-2r}{
    m_2^2}\right)^{{}^1m}\Biggl\{ C^{n_2}_{2{}^1m}\left|\vec{N}(2)+\delta_{ij,11} +({}^1m-1)
    \delta_{ij,22}+\delta_{s,12}-2{}^1m\delta_{s,2} \rangle_V\right.\nonumber\\
   &\hspace{-1em}&\hspace{-1em} -\Bigl[C^{n_2}_{2{}^1m}
     (h^2-h^1+2p_{12})+
     C^{n_2}_{2{}^1m+1}\Bigr]\left|\vec{N}(2)\hspace{-0.1em}+\hspace{-0.1em}
     \delta_{ij,12}\hspace{-0.1em}+\hspace{-0.1em}({}^1m-1)\delta_{ij,22}\hspace{-0.1em}
     -\hspace{-0.1em}
     2{}^1m\delta_{s,2}\rangle_V\right.\nonumber\\
     &\hspace{-1em}&\hspace{-1em} + p_{12}C^{n_2}_{2{}^1m}
     (h^2-h^1+p_{12}-1)\left|\vec{N}(2)+{}^1m\delta_{ij,22}-\delta_{s,12}-
     2{}^1m\delta_{s,2}\rangle_V\right.
 \Biggr\}\nonumber\\
&\hspace{-1em}&\hspace{-1em} -
\sum\limits^{\left[n_2/2\right]}_{{}^{1}m=0}
  \sum\limits^{\left[n_2/2-({}^{1}m + 1)\right]}_{{}^1l=0}
 \hspace{-0.3em}\left(\frac{-2r}{
    m_2^2}\right)^{{}^{1}m+{}^{1}l +1}\hspace{-0.7em}
    C^{n_2}_{2{}^{1}m+1}\Biggl\{\Bigl[
    C^{n_2-2{}^{1}m-1}_{2{}^1l+1}
    \bigl(h^2-h^1+ 2p_{12}+n_2\nonumber\\
&\hspace{-1em}&\hspace{-1em} - 2({}^{1}m
+{}^1l+1)\bigr)-C^{n_2-2{}^{1}m -1 }_{2{}^1l+2}
    \Bigr]
   \left|\vec{N}(2)+\delta_{ij,12}+({}^1m+{}^1l)\delta_{ij,22}-2({}^{1}m\right.
   \nonumber\\
&\hspace{-1em}&\hspace{-1em} \left.
+{}^{1}l+1)\delta_{s,2}\rangle_V\right.
-p_{12}C^{n_2-2{}^{1}m-1}_{2{}^1l+1}
    \bigl(h^2-h^1+ n_2-2({}^{1}m  +{}^1l+1)+p_{12}-1\bigr)\nonumber\\
    &\hspace{-1em}&\hspace{-1em}\times \left|\vec{N}(2)+({}^1m+{}^1l+1)\delta_{ij,22}-\delta_{s,12}-
2({}^{1}m +{}^{1}l+1)\delta_{s,2}\rangle_V\right.
\nonumber\\
  &\hspace{-1em}&\hspace{-1em}-
C^{n_2-2{}^{1}m-1}_{2{}^1l+1}\left|\vec{N}(2)+\delta_{ij,11}+({}^1m+{}^1l)\delta_{ij,22}+\delta_{s,12}-
2({}^{1}m +{}^{1}l+1)\delta_{s,2}\rangle_V\right.\nonumber\\
  &\hspace{-1em}&\hspace{-1em}
+
\frac{m_1}{m_2}C^{n_2-2{}^{1}m-1}_{2{}^1l+2}\left|\vec{N}(2)\hspace{-0.2em}
+\hspace{-0.2em}
({}^1m+{}^1l+1)\delta_{ij,22}\hspace{-0.2em}+\hspace{-0.2em}
\delta_{s,1}\hspace{-0.2em}-\hspace{-0.2em} 2({}^{1}m
+{}^{1}l+\textstyle\frac{3}{2})\delta_{s,2}\rangle_V\right.
      \Biggr\}.
\end{eqnarray}

The obtained result has obvious  consequences, first, in case of
reducing of $\mathcal{A}'(Y(2), AdS_d)$ to the quadratic  algebra
$\mathcal{A}'(Y(1), AdS_d)$, given in Ref. \cite{BurdikNavratilPasnev, BKL}
for the
vanishing components $n_{l2}=n_2=p_{12}=0, l=1,2$ of an arbitrary
VM vector $\left|\vec{N}(1)\rangle_V\right. \equiv \left|n_{11},
0,0,n_1,0,0\rangle_V\right.$. Second,  for reducing the AdS space
to Minkowski $R^{1,d-1}$ space when the non-linear algebra
$\mathcal{A}'(Y(2), AdS_d)$ for $r=0$ turns to the Lie algebra
$\left(T^{\prime 2} \oplus T^{\prime 2*}\oplus l'_0\right)(0) +
\hspace{-1em} \supset
  sp(4)$ for $k=2$ in the Eqs.(\ref{identalgaux}). For former case
   we derive
  known Verma module \cite{BKL} from results of Theorem~1 whereas for latter one
  we will get new Verma module realization (see \cite{BurdikReshetnyak}) for above Lie algebra being
  different from one for $sp(4)$ in \cite{flatbos} and in \cite{BuchbinderReshetnyak}
 for $k=2$.

\subsection{Fock space realization of
$\mathcal{A}'(Y(2), AdS_d)$}\label{oscVMY2b}

In this section, we will find on a base of constructed Verma
module the realization of $\mathcal{A}'(Y(2), AdS_d)$ in formal
power series  in degrees of creation and annihilation operators
 $(B_a,B^+_a) = \bigl((b_i,
b_{ij}, d_{12}), (b_i^+, b_{ij}^+, d_{12}^+)\bigr)$ in
$\mathcal{H}'$ whose number coincides to ones of second-class
constraints among $o'_I$, i.e. with
$\dim(\mathcal{E}^-_2\oplus\mathcal{E}^+_2)$.  It  is solved
following the results of \cite{Liealgebra} and algorithms
suggested in \cite{BurLeites} initially elaborated for a simple
Lie algebra, then enlarged  to a non-linear quadratic algebra
$\mathcal{A}(Y(1), AdS_d)$ Ref.~\cite{BurdikNavratilPasnev}. To
this end, we make use of the mapping for an arbitrary basis vector
of the Verma module and one $\left|\vec{n}_{ij}, \vec{n}_s\rangle
\right.$ of  new Fock space $\mathcal{H}'$,
\begin{eqnarray}\label{map}
   & \left|\vec{n}_{ij}, \vec{n}_s\rangle_V \right.
    \longleftrightarrow \left|\vec{n}_{ij}, \vec{n}_s\rangle \right.
 = \prod_{i\leq j}^2\bigl(b^+_{ij}\bigr){}^{n_{ij}}\prod_{l=1}^2\bigl(b^{+}_l\bigr){}^{
 n_{l}}\bigl(d^{+}_{12}\bigr){}^{p_{12}}|0\rangle\,, &\\
&  \mathrm{for}\ b_{ij}|0\rangle = b_l|0\rangle = d_{12} |0\rangle
= 0,
 \quad \Bigl(\left|\vec{N}(2)\rangle \right.{}_{[\vec{N}=\vec{0}]} \equiv |0\rangle
 \Bigr).&\nonumber
\end{eqnarray}
Here, vector $\left|\vec{n}_{ij}, \vec{n}_s\rangle \right.$, for
non-negative integers $n_{ij}, n_l, p_{12}$ are the basis vectors
of a Fock space $\mathcal{H}'$ generated by 6 pair of bosonic
$(B,B^+)$ operators, being the basis elements of the Heisenberg
 algebra $A_{6}$, with the standard
(only non-vanishing) commutation relations
\begin{equation}\label{commrelations}
 [B_a,B^+_b] =
 \delta_{ab} \Longleftrightarrow [b_{ij},b^+_{lm}]=\delta_{il,jm},\ [b_{i},b^+_{m}]=\delta_{im},\ [d_{12},d^+_{12}]=1 .
\end{equation}
Omitting the peculiarities of the
 correspondence among the  Verma module vectors (especially one of
$\left|\hat{A}_{\vec{N}(2)}\rangle_V\right.$ for $\vec{N}(2)=(0,...,0,n_2)$
(\ref{Antot}) and
Fock space $\mathcal{H}'$ vector
$\left|\vec{0}_{ij},0,0,n_2\rangle\right.$  (see \cite{BurdikReshetnyak}
for details) we formulate our basic result
as the

\vspace{1ex} \noindent \textbf{\emph{{Theorem 2.}}} The oscillator
realization of the non-linear  algebra
$\mathcal{A}'(Y(2), AdS_d)$ over Heisenberg algebra $A_6$ exists
in terms of formal power series in degrees of creation and
annihilation operators, is given by the relations
(\ref{l'+ijF})--(\ref{t+totF}), (\ref{l2'+F}),
(\ref{l0fin})--(\ref{l22fin})   and expressed with help of
\emph{primary  block-operator} $\widehat{t}'_{12}$
(\ref{hattosc}),  and \emph{derived block-operators},
$\widehat{t}^{\prime +}_{12}, \widehat{l}^{\prime }_{0} ,
\widehat{l}^{\prime }_{m}, \widehat{l}^{\prime }_{m2}, m=1,2$
(\ref{t'+F}), (\ref{l0oscnn2})--(\ref{l22oscnn2}) as follows.

First, for the trivial negative root vectors, we have
\begin{eqnarray}
 && l^{\prime  +}_1  =  m_1b_1^+
  \,,\quad l^{\prime+}_{ij}  =  b_{ij}^+\,,
  \quad g_0^{\prime i} =
 2b_{ii}^+b_{ii}+b_{12}^+b_{12} + (-1)^id^+_{12}d_{12} + b^+_ib_i+ h^i .
 \label{l'+ijF}
\end{eqnarray}
Second, for the operator ${t}'_{12}(B,B^+)$ and \emph{primary
block-operator} $\widehat{t}'_{12}(B,B^+)$   we obtain,
 {\begin{eqnarray}
   {t}'_{12} &=&
   \left(h^1-h^2 - b^+_2b_2 - d^+_{12}d_{12}\right)d_{12} -
   b_{11}^+b_{12}
   -2b_{12}^+b_{22} +\widehat{t}'_{12} \,,
 \label{t'Lf}  \\
\widehat{t}'_{12} & =&
\sum_{k=0}\Biggl[\sum_{{}^1m=0}\hspace{-0.2em}
   \sum_{{}^1l=0} ... \sum_{{}^km=0}\hspace{-0.2em}\sum_{{}^kl=0}
   (-1)^k\hspace{-0.2em}
   \left(\hspace{-0.2em}\frac{-2r}{m_2^2}\hspace{-0.2em}\right)^{\hspace{-0.2em}\sum_{i=1}^k({}^im+{}^il)+k}
   \hspace{-0.2em}\prod_{i=1}^k
   \frac{1}{(2{}^im+1)!}\frac{1}{(2{}^il+1)!}\nonumber
\label{hattosc}     \\
   && \times \bigl(b_{22}^+\bigr)^{\sum_{i=1}^k({}^im+{}^il)+k}\Biggl\{
   b_2^+d_{12}b_2 -\frac{m_1}{m_2}\sum_{m=0}\left(\frac{-2r}{m_2^2}\right)^m
    \frac{(b_{22}^+)^m}{(2m+1)!}b_1^+
    b_2^{2m+1}\nonumber \\
   && +\sum_{m=1}\left(\frac{-2r}{m_2^2}\right)^m
   (b_{22}^+)^{m-1}\Bigl[b_{12}^+
   \Bigl\{
    \frac{h^2-h^1+2d^+_{12}d_{12}}{(2m)!}+\frac{b^+_2b_2}{(2m+1)!}\Bigr\}
   \nonumber \\
  &&  -
\frac{b_{11}^+d^+_{12}}{(2m)!} -
\frac{b_{22}^+}{(2m)!}(h^2-h^1+d^+_{12}d_{12})d_{12}\Bigr]b^{2m}_2 \nonumber \\
     &&   - \sum_{m=0}\sum_{l=0}\left(\frac{-2r}{m_2^2}\right)^{m+l+1}
   \frac{1}{(2m+1)!}(b_{22}^+)^{m+l}\Bigl[ b_{12}^+
   \Bigl\{\frac{h^2-h^1+2d^+_{12}d_{12}+b_2^+b_2}{(2l+1)!}\nonumber\\
 && -
   \frac{b_2^+b_2}{(2l+2)!}
   \Bigr\}
   -\frac{b_{11}^+d^+_{12}}{(2l+1)!}  -\frac{b_{22}^+}{(2l+1)!}(h^2-h^1+
   b_2^+b_2+d^+_{12}d_{12})d_{12} \nonumber\\
      &&  +
   \frac{m_1}{m_2}\frac{b_{22}^+}{(2l+2)!}b_1^+b_2
   \Bigr]  b_2^{2(m+l+1)}
   \Biggr\}(b_2)^{2(\sum_{i=1}^k({}^im+{}^il)+k)}\Biggr]\footnotemark.
\end{eqnarray}}\footnotetext{one should be noted that in (\ref{hattosc})
 for $k=0$ there are no doubled sums and the products
$\prod_{i=1}^0...$ is equal to  $1$ and the only term
$b_2^+d_{12}b_2$ survive}Third, for the operators
$t^{\prime+}_{12}, l^{\prime+}_2$ and derived block-operator
$\widehat{t}^{\prime+}_{12}$ we have,
\begin{eqnarray}
 t^{\prime+}_{12} &=&
 - 2 b_{12}^+b_{11}-b_{22}^+b_{12}
 +\widehat{t}^{\prime+}_{12}, \label{t+totF} \\
   \widehat{t}^{\prime+}_{12}  & = &
 \sum_{m=0}\left(\frac{-2r}{m_1^2}\right)^m(b_{11}^+)^m\Bigl\{
\frac{d^+_{12}}{(2m)!}
-\frac{m_2}{m_1}\frac{b_2^+b_1}{(2m+1)!}\Bigr\}b_1^{2m}\nonumber\\
 && \hspace{1.0em}
-\hspace{-0.2em}\sum_{m=1}\hspace{-0.2em}\left(\hspace{-0.2em}
\frac{-2r}{m_1^2}\hspace{-0.2em}\right)^m\hspace{-0.2em}
(b_{11}^+)^{m-1}\Biggl\{
b_{12}^+\hspace{-0.2em}\left[\frac{(h^2\hspace{-0.2em}-h^1\hspace{-0.2em}+2d^+_{12}d_{12}\hspace{-0.2em}+b^+_2b_2)}{(2m)!}
\hspace{-0.2em}-\frac{b_1^+b_1}{(2m+1)!}\hspace{-0.2em}\right] \nonumber\\
 && \hspace{1.0em} +b_{22}^+\hspace{-0.2em}
\frac{(h^1\hspace{-0.2em}-h^2\hspace{-0.2em}-d^+_{12}d_{12}
\hspace{-0.2em}-b^+_2b_2)}{(2m)!}d_{12} + \frac{b_{22}^+}{ (2m)!}\
\widehat{t}'_{12}\ \Biggr\}b_1^{2m} , \label{t'+F}\\
%
%
   l^{\prime+}_2 &= &
m_1\sum_{m=0}\hspace{-0.2em}\left(\hspace{-0.2em}\frac{-2r}{m_1^2}\hspace{-0.2em}
\right)^{m+1}\hspace{-0.5em}(b_{11}^+)^{m}\Biggl\{
b_{12}^+\left[\frac{(h^2-h^1+2d^+_{12}d_{12}+b^+_2b_2)}{(2m+1)!}
-\frac{b_1^+b_1}{(2m+2)!}\right]  \nonumber\\
 &&
 + b_{22}^+
\frac{(h^1-h^2-d^+_{12}d_{12}-b^+_2b_2)}{(2m+1)!}d_{12}-
b_{11}^+\frac{d^+_{12}}{(2m+1)!}\Biggr\}b_1^{2m+1}
\nonumber\\
&& +m_2\sum_{m=0}\left(\frac{-2r}{m_1^2}\right)^m
\frac{(b_{11}^+)^m b_2^+}{(2m)!} b_1^{2m}
+m_1\sum_{m=0}\left(\frac{-2r}{m_1^2}\right)^{m+1}\frac{(b_{11}^+)^{m}
b_{22}^+}{(2m+1)!}\ \widehat{t}'_{12}\ b_{1}^{2m+1}. \label{l2'+F}
\end{eqnarray}
Fourth, for the Cartan-like vector ${l^{\prime }_0}$
 the representation holds,
 \begin{eqnarray}
   {l^{\prime  }_0} &=& \widehat{l}^{\prime  }_0 +
\frac{m_1}{2}\displaystyle\sum\limits_{m=0}\left(
\frac{-8r}{m_1^2}\right)^{m+1}\frac{(b_{11}^+)^m}{(2m+1)!}\Biggl\{
b_{12}^+ \ \widehat{l}^{\prime }_2\  +\ b_{11}^+ \
\widehat{l}^{\prime }_1\nonumber \\
\hspace{-1.0em} && \quad - \frac{m_2}{2}
[\widehat{t}^{\prime}_{12} +(h^1-h^2-b^{2+}b_2 -
d^+_{12}d_{12})d_{12}]\ b_2^+ \Biggr\}b_1^{2m+1} \nonumber
\\
&&  - r \displaystyle\sum\limits_{m=0}\left(
\frac{-8r}{m_1^2}\right)^{m}(b_{11}^+)^m b_1^+\left\{
\frac{2(h^1-d^+_{12}d_{12})-3}{(2m+1)!} +
\frac{2b_1^+b_1}{(2m+2)!}\right\}b_1^{2m+1}\nonumber \\
 &&  +
\frac{1}{2}\hspace{-0.2em}\displaystyle\sum\limits_{m=0}\hspace{-0.2em}
\left(\hspace{-0.2em}
\frac{-8r}{m_1^2}\hspace{-0.2em}\right)^{m+1}\hspace{-1.0em}
\frac{(b_{11}^+)^{m}}{ (2m+2)!}\Biggl\{b_{11}^+\
\widehat{l}^{\prime }_0\
 \hspace{-0.2em}-rb_{11}^+\Bigl([h^1\hspace{-0.2em} - d^+_{12}d_{12}][h^1 \hspace{-0.2em}
 - d^+_{12}d_{12} \hspace{-0.2em}- 2]\nonumber
\\
 &&  \quad - h^2 - d^+_{12}d_{12} - b^+_2b_2 +
2d^+_{12}(h^1 - h^2 - b^+_2b_2 - d^+_{12}d_{12})d_{12}\Bigr)
\nonumber
\\
 &&  \quad -2r b_{11}^+\widehat{t}'_{12}\ d^+_{12}
+ 8r b_{11}^+ b_{12}^+ \widehat{l}^{\prime}_{12} \ +4r
(b_{12}^+)^2\ \widehat{l}^{\prime}_{22} + 2rb_{12}^+ \Bigl[\
(h^1\hspace{-0.2em} - h^2
 \nonumber\\
 && \quad  - b^+_2b_2 - d^+_{12}d_{12})d_{12}+
\widehat{t}'_{12}\Bigr] \ \Bigl[ h^2 + b^+_2b_2 + d^+_{12}d_{12} -
2\Bigr]\Biggr\}b_1^{2m+2}
 \nonumber\\
 &&  +
\frac{r}{2}\displaystyle\sum\limits_{m=0}\left(
\frac{-8r}{m_1^2}\right)^{m+1}\frac{(b_{11}^+)^m}{(2m+2)!}b_{22}^+\
\Biggl\{\Bigl[(h^1-h^2-b^+_2b_2-d^+_{12}d_{12}) d_{12} +
\widehat{t'}_{12}
   \Bigr]\nonumber\\
&& \qquad
\times(h^1\hspace{-0.2em}-h^2\hspace{-0.2em}-b^+_2b_2\hspace{-0.2em}-d^+_{12}d_{12})
d_{12}\hspace{-0.2em} + t'_{12}\widehat{t}'_{12} \Biggr\}\
b_1^{2m+2}
 \label{l0fin}.
\end{eqnarray}
Fifth, the operators $l^{\prime }_m$,  $l^{\prime }_{lm}$,
$l,m=1,2, l\leq m$, read as follows,
\begin{eqnarray}
l^{\prime }_1 & =& -m_1b^+_1b_{11}
 -\frac{1}{2}{l}^{\prime 2+}b_{12}  +\sum\limits_{m=0}
\left(\frac{-8r}{m_1^2}\right)^{m}\frac{(b_{11}^+)^m}{(2m)!}
\widehat{l}^{\prime 1}b_1^{2m}  \nonumber\\
&& + \frac{1}{m_1}\displaystyle\sum\limits_{m=0}\left(
\frac{-8r}{m_1^2}\right)^{m}{\frac{(b_{11}^+)^m}{(2m+1)!}}
\Biggl\{\widehat{l}^{\prime }_0  -r
\Bigl[(h^1-d^+_{12}d_{12})(h^1-d^+_{12}d_{12} -2) -
h^2\nonumber\\
&&  -d^+_{12}d_{12}-b^+_{2}b_{2}\Bigr] -
rd^+_{12}\Bigl[h^1-h^2-b_2^+b_2 -d^+_{12}d_{12}\Bigr] d_{12}
 - r\ \widehat{t}'_{12}\ d^+_{12} \Biggr\}b_1^{2m+1}\nonumber \\
&& - 4{m_1}\displaystyle\sum\limits_{m=0} \left(
\frac{-2r}{m_1^2}\right)^{m+1}{\frac{(b_{11}^+)^m
b_{12}^+}{(2m+1)!}}
\textstyle(4^m-\frac{1}{2})\ \widehat{l}'_{12}\ b_1^{2m+1}\nonumber\\
&&
 + \frac{m_1}{2}\displaystyle\sum\limits_{m=1}\left(
\frac{-8r}{m_1^2}\right)^{m}(b_{11}^+)^{m-1}b_1^+
\left\{\frac{1}{2}\frac{(2(h^1-d^+_{12}d_{12})-1)}{(2m)!} +
\frac{b_1^+b_1}{(2m+1)!}\right\}b_1^{2m} \nonumber\\
&& + \frac{1}{2}\displaystyle\sum\limits_{m=1}\left(
\frac{-2r}{m_1^2}\right)^{m}(4^m-1)
{\frac{(b_{11}^+)^{m-1}}{(2m)!}}\Bigl\{2b_{12}^+
\widehat{l}^{\prime 2}   - m_1b_1^+\nonumber\\
&& - m_2 (\widehat{t}^{\prime }_{12}+
h^1-h^2-b_2^+b_2-d^+_{12}d_{12})d_{12}b_2^+ \Bigr\}b_1^{2m}
\nonumber\\
&& +\frac{m_1}{2}\displaystyle\sum\limits_{m=1}\left(
\frac{-2r}{m_1^2}\right)^{m+1}(4^m-1)
  {\frac{(b_{11}^+)^{m-1}}{(2m+1)!}}
  \Biggl\{ b_{11}^+\ \widehat{t}'_{12}\ d^+_{12} -
  4(b_{12}^+)^2\widehat{l}^{\prime }_{22} \nonumber\\
  && \qquad -2 b_{12}^+
  \Bigl[(h^1-h^2- b_2^+b_2 - d^+_{12}d_{12})d_{12}  +
  \widehat{t}'_{12} \Bigr]
  (h^2 + b_2^+b_2 + d^+_{12}d_{12} - 2)\nonumber\\
&& \qquad+ b_{11}^+d^+_{12}(h^1-h^2-b_2^+b_2-d^+_{12}d_{12})d_{12}
    \Biggr\}b_1^{2m+1}\nonumber\\
&& -
\frac{m_1}{2}\displaystyle\sum\limits_{m=1}\hspace{-0.2em}\left(
\frac{-2r}{m_1^2}\hspace{-0.2em}\right)^{m+1}\hspace{-1.0em}(4^m-1)
  {\frac{(b_{11}^+)^{m-1} }{(2m+1)!}}b_{22}^+
  \Biggl\{\Bigl[(h^1-h^2-b^+_2b_2-d^+_{12}d_{12}) d_{12} \nonumber\\
&&\qquad  + \widehat{t}'_{12}
   \Bigr](h^1\hspace{-0.2em}-h^2\hspace{-0.2em}-b^+_2b_2\hspace{-0.2em}-d^+_{12}d_{12})
d_{12}\hspace{-0.2em} + {t}'_{12}\widehat{t}'_{12}
\Biggr\}b_1^{2m+1}, \label{l1'F}
\\
{l^{\prime }_2}
 &= &
\displaystyle\sum\limits_{m=0}\hspace{-0.2em}\left(
\hspace{-0.2em}\frac{-2r}{m_1^2}\hspace{-0.2em}\right)^{m}\hspace{-0.2em}\frac{(b_{11}^+)^m}{(2m+2)!}
\widehat{l}^{\prime }_2\
(b_1^+)^2b_1^{2m+2}\hspace{-0.2em}-2{m_1}\hspace{-0.2em}\displaystyle\sum\limits_{m=0}
 \hspace{-0.2em}\left( \hspace{-0.2em}\frac{-2r}{m_1^2}\hspace{-0.2em}\right)^{m+1}
\frac{(b_{11}^+)^{m}}{(2m+1)!}\sum_kb_{1k}^+ \widehat{l}^{\prime
}_{k2}
b_1^{2m+1}\nonumber\\
&& -2m_1 \displaystyle\sum\limits_{m=0}\left(
\frac{-2r}{m_1^2}\right)^{m+1}
\hspace{-0.2em}\frac{(b_{11}^+)^m}{4} \widehat{t}'_{12}
\hspace{-0.2em}\left\{
  \frac{(h^1+h^2 + b_2^+b_2 -
2)}{(2m+1)!} + \frac{b_1^+b_1}{(2m+2)!}\right\}
 b_1^{2m+1}\nonumber\\
&&-\frac{m_1}{2}\displaystyle\sum\limits_{m=0} \left(
\frac{-2r}{m_1^2}\right)^{m+1}(b_{11}^+)^{m}\Biggl\{
\frac{(h^1+h^2+b_2^+b_2 - 2)}{(2m+1)!} +
\frac{b_1^+b_1}{(2m+2)!}\Biggr\}\times\nonumber\\
&& \quad \times(h^1-h^2-b_2^+b_2-d_{12}^+d_{12})d_{12}b_1^{2m+1} -
\frac{m_1}{2}b^+_1b_{12} -  {l}^{\prime+}_2\ b_{22}, \label{l2fin}
\\
{l'_{11}}  &=&
  ({b}^+_{11}b_{11}+{b}^+_{12}{b}_{12}+b^+_1b_1 -d^+_{12} d_{12}
   + h^1)b_{11}
    + \frac{b_{22}^+}{4}b_{12}^2
 - \frac{1}{2}\widehat{t}^{\prime
+}_{12}b_{12} \nonumber\\
&& + \frac{1}{2}\sum_{m=0}  \left(\frac{-8r}{m_1^2}
\right)^{m+1}\frac{(b_{11}^+)^{m}}{(2m+2)!}\Biggl\{\frac{1}{4r}
\widehat{l}_0'\ + b_{12}^+\ \widehat{l}^{\prime }_{12} -
\frac{1}{4}\Bigl[d^+_{12}(2-h^1-h^2-b^+_2b_2)d_{12} \nonumber\\
&&\quad + h^1(h^1-2) - h^2-b^+_2b_2+\widehat{t}'_{12} d^+_{12}  \Bigr]
\Biggr\} b_1^{2m+2} \nonumber\\
&& - \frac{1}{2}\sum_{m=1} \left(\frac{-8r}{m_1^2} \right)^{m}
(b_{11}^+)^{m-1}b_1^+\left\{\frac{h^1-d^+_{12}d_{12}-\frac{1}{2}}{(2m+1)!}
+\frac{b_1^+b_1}{(2m+2)!}
\right\}b_1^{2m+1}\nonumber\\
&&-\frac{1}{m_1}\sum_{m=0}\Biggl\{\left(\frac{-8r}{m_1^2}
\right)^{m}\frac{(b_{11}^+)^{m}}{(2m+1)!}\ \widehat{l}^{\prime
}_1\
  - \left(\frac{-2r}{m_1^2}
\right)^{m}(4^m-1) \frac{(b_{11}^+)^{m-1}b_{12}^+}{(2m+1)!}\
\widehat{l}^{\prime }_2\Biggr\}\ b_{1}^{2m+1}
\nonumber\\
&& + \frac{1}{2} \sum_{m=0}\left(\frac{-2r}{m_1^2}
\right)^{m}(4^m-1)
(b_{11}^+)^{m-1}\frac{b_1^+b_1}{(2m+1)!}b_{1}^{2m}
\nonumber\\
&& +\frac{m_2}{2m_1}\hspace{-0.2em}
\sum_{m=0}\hspace{-0.2em}\left(\hspace{-0.2em}\frac{-2r}{m_1^2}\hspace{-0.2em}
\right)^{m}\hspace{-0.5em}(4^m-1)
\frac{(b_{11}^+)^{m-1}}{(2m+1)!}\ \Bigl\{
{t}^{\prime}_{12}\hspace{-0.2em}  + \sum_l lb^+_{1l}b_{l2}\Bigr\}
b_2^+ b_{1}^{2m+1}
\nonumber\\
&& -\frac{1}{2}
\hspace{-0.2em}\sum_{m=0}\hspace{-0.2em}\left(\hspace{-0.2em}\frac{-2r}{m_1^2}
\hspace{-0.2em}\right)^{m+1}\hspace{-0.7em}(4^m-1)
\frac{(b_{11}^+)^{m-1}}{(2m+2)!}\left[b_{11}^+d^+_{12}\hspace{-0.2em}
-2b_{12}^+( h^2+b_2^+b_2+d_{12}^+d_{12}-1)\right]\times
\nonumber\\
&& \qquad \times\Bigl(h^1-h^2-d_{12}^+d_{12}-b^+_2b_2\Bigr)d_{12}
b_{1}^{2m+2}
\nonumber\\
&&-\frac{1}{2} \sum_{m=0}\left(\frac{-2r}{m_1^2}
\right)^{m+1}(4^m-1) \frac{(b_{11}^+)^{m-1}}{(2m+2)!} \Bigl\{\
b_{11}^+\ \widehat{t}'_{12}\ d^+_{12} -4b_{11}^+b_{12}^+\
\widehat{l}^{\prime }_{12}\ \nonumber\\
&&\qquad -4(b_{12}^+)^2\ \widehat{l}^{\prime }_{22}\
 -2 b_{12}^+ \ \widehat{t}'_{12}\ (h^2+b^+_2b_2+d^+_{12}d_{12}-2)\Bigr\}b_{1}^{2m+2}
 \nonumber\\
&& + \frac{1}{2} \sum_{m=0}\left(\frac{-2r}{m_1^2}
\right)^{m+1}(4^m-1) \frac{(b_{11}^+)^{m-1}}{(2m+2)!} b_{22}^+ \
\left\{\Bigl[(h^1-h^2-b^+_2b_2-d^+_{12}d_{12})
d_{12}+\widehat{t}'_{12}
   \Bigr] \times\right.\nonumber\\
&&\qquad \left.\times \bigl(h^1- h^2 -b^+_2b_2
-d^+_{12}d_{12}\bigr) d_{12} + {t}'_{12}\widehat{t}'_{12} \right\}
b_{1}^{2m+2},
 \label{l11fin}
\\%
 l'_{12} & = &
\frac{1}{4}\Bigl( b_{12}^+b_{12}+
 \displaystyle\sum_{m}(2b_{mm}^+b_{mm}+b_m^+b_m+h^m)\Bigr)
 b_{12}+ b_{12}^+b_{11}b_{22}
-\frac{1}{2} \left(\widehat{t}^{\prime +}_{12}\ b_{22} +
\widehat{t}^{\prime }_{12}\ b_{11}\right)
 \nonumber \\
 && + \frac{1}{2} (h^2-h^1 + b_2^+b_2 + d^+_{12}d_{12} )b_{11}d_{12}
    -
\frac{1}{2m_1}\hspace{-0.2em}\sum_{m=0}
\hspace{-0.2em}\left(\hspace{-0.2em}\frac{-2r}{m_1^2}\hspace{-0.2em}
\right)^{m}\hspace{-0.3em}\frac{(b_{11}^+)^m}{(2m+1)!}
\widehat{l}^{\prime
}_2b_1^{2m+1}\nonumber\\
&& + \sum_{m=0}  \left(\frac{-2r}{m_1^2}
\right)^{m}\frac{(b_{11}^+)^m}{(2m)!} \widehat{l}^{\prime
}_{12}b_1^{2m}+ \sum_{m=1}\left(\frac{-2r}{m_1^2} \right)^{m}
\frac{(b_{11}^+)^{m-1}b_{12}^+}{(2m)!}\ \widehat{l}^{\prime
 }_{22}b_1^{2m}\nonumber\\
&& + \frac{1}{4}\sum_{m=1}\left(\frac{-2r}{m_1^2} \right)^{m}
(b_{11}^+)^{m-1} \Biggl\{\widehat{t}'_{12}\
\left[\frac{(h^1+h^2+b_2^+b_2-2)}{(2m)!}
+\frac{b_1^+b_1}{(2m+1)!}\right] \nonumber \\
&& +\left[\frac{h^1+h^2+b_2^+b_2-2}{(2m)!}
+\frac{b_1^+b_1}{(2m+1)!} \right]
(h^1-h^2-b^+_2b_2-d_{12}^+d_{12})d_{12} \Biggr\} b_1^{2m},
 \label{l12fin}\\
l^{\prime }_{22}  & = & \widehat{l}^{\prime }_{22}\ + \bigl(
b^+_{12}b_{12} + d^+_{12}d_{12} + b^+_2b_2 + b^+_{22}b_{22}+
h^2\bigr)b_{22}  +  \frac{1}{4}b_{11}^+b_{12}^2
   -\frac{1}{2}\ \widehat{t}'_{12}\ b_{12}
 \nonumber\\
  &&+\frac{1}{2}( h^2 - h^1+d^+_{12}d_{12} +b^+_2b_2)d_{12}b_{12}
. \label{l22fin}
\end{eqnarray}
In the Eqs.(\ref{l0fin})--(\ref{l22fin}), the  \emph{derived block-operators} $\widehat{l}^{\prime}_0$, $\widehat{l}^{\prime}_m$,
$\widehat{l}^{\prime}_{m2}$, for $m=1,2$
 are written  as
follows,\footnote{they are directly determined through theirs
action  on vector
$\left|\vec{0}_{ij},0,p_{12},n_2\rangle_V\right.$, see \cite{BurdikReshetnyak} for details}
\begin{eqnarray}
{\widehat{l}^{\prime  }}_0  & = & m_0^2
-r\sum_{m=0}\left(\frac{-8r}{m_2^2} \right)^{m}(b_{22}^+)^{m}
\Biggl\{b_2^+\left[\frac{1}{(2m+1)!}(2h^2+2d^+_{12}d_{12} -
1)\right. \nonumber
\\
&&\quad \left.+ \frac{2b_2^+b_2}{(2m+2)!}\right]b_2
-2\frac{m_1}{m_2}\frac{b_1^+d^+_{12}}{(2m+1)!}b_2\Biggr\}b_2^{2m}\nonumber\\
\hspace{-1.0em} && +\frac{1}{2} \sum_{m=0}\left(\frac{-8r}{m_2^2}
\right)^{m+1}
\frac{(b_{22}^+)^{m}}{(2m+2)!}\Bigl\{b_{22}^+\Bigl(m_0^2-
r\bigl[h^2(h^2-4)+h^1-(d^+_{12})^2d_{12}^2\nonumber\\
\hspace{-1.0em} && \quad   +2d^+_{12}d_{12}(h^1-2)\bigr]\Bigr)+
2rb_{12}^+d^+_{12} (h^1 - d^+_{12}d_{12} - 2) +
rb_{11}^+(d^+_{12})^2\Bigr\}b_2^{2m+2}\nonumber\\
\hspace{-1.0em} && +2r\sum_{m=0}\sum_{l=0}\left(\frac{-8r}{m_2^2}
\right)^{m} \left(\frac{-2r}{m_2^2}
\right)^{l+1}\frac{(b_{22}^+)^m}{(2m+1)!}
\frac{(b_{22}^+)^{l+1}}{(2l+1)!} \ \widehat{t}'_{12}\ d^+_{12}
b_2^{2(m+l+1)}\nonumber\\
 && +2r\sum_{m=0}\sum_{l=0}\left(
\frac{-8r}{m_2^2}\right)^{m} \left(\frac{-2r}{m_2^2}
\right)^{l+1}\frac{(b_{22}^+)^{m+l}}{(2m+1)!}\Biggl\{
b_{12}^+d^+_{12}\hspace{-0.2em}\Biggl[\hspace{-0.2em}\frac{[h^2\hspace{-0.2em}-h^1
\hspace{-0.2em}+2d^+_{12}d_{12}\hspace{-0.2em}+b^+_2b_2\hspace{-0.2em}
+2]}{(2l+1)!}  \nonumber\\
\hspace{-1.0em} &&
 - \frac{b_2^+ b_2}{(2l+2)!}\Biggr] -
\frac{b_{11}^+(d^+_{12})^2}{(2l+1)!} +
\frac{b_{22}^+(d^+_{12}d_{12} +1) }{(2l+1)!}  [h^1 - h^2 -
b_2^+b_2 - d^+_{12}d_{12}]\nonumber\\
\hspace{-1.0em} && \quad +
\frac{m_1}{m_2}\frac{b_{22}^+b_1^+d^+_{12}}{(2l+2)!}
b_2\Biggr\}b_2^{2(m+l+1)}, \label{l0oscnn2}
\\
 \widehat{l}^{\prime  }_1  &=&
-\frac{m_2}{2}\sum_{m=1}\left( \frac{-2r}{m_2^2} \right)^{m+1}
(b_{22}^+)^{m} d^+_{12}\left\{\frac{(h^1+h^2-2)}{(2m+1)!}+
\frac{b_2^+b_2}{(2m+2)!}\right\}b_2^{2m+1} ,
\label{l1oscnn2}\\
\widehat{l}^{\prime }_2 & =& - \widehat{l}^{\prime }_1 d_{12}
-\sum_{m=0} \left(\frac{-8r}{m_2^2}
\right)^{m}\frac{(b_{22}^+)^m}{(2m+1)!}
\frac{[m_0^2-rh^2(h^2-3)]}{m_2}b_2^{2m+1} \nonumber\\
 &&  + \frac{m_2}{4} \sum_{m=1}
\left(\frac{-8r}{m_2^2}
\right)^{m}(b_{22}^+)^{m-1}b_2^+\left\{\frac{(2h^2-1)}{(2m)!} +
2\frac{b_2^+b_2}{(2m+1)!}\right\}b_2^{2m}\nonumber\\
 &&
 +
\frac{m_2}{2}\sum_{m=1}\left(\hspace{-0.2em}\frac{-2r}{m_2^2}
\right)^{m+1} (b_{22}^+)^m
d^+_{12}\left\{\frac{(h^1+h^2-2)}{(2m+1)!}+
\frac{b_2^+b_2}{(2m+2)!}\right\}d_{12} b_2^{2m+1} \nonumber
\\
 &&  -  \frac{1}{2}\sum_{m=0}
\left(\frac{-2r}{m_2^2} \right)^{m}(4^m-1)(b_{22}^+)^{m-1}
d^+_{12}\left\{\frac{{m_1}b_1^+}{(2m)!}-\frac{{m_2}b_2^+d_{12}}{(2m)!}
\right\}
b_2^{2m} \nonumber\\
 &&  + \frac{m_2}{2} \sum_{m=0}
\left(\frac{-2r}{m_2^2}
\right)^{m+1}(4^m-1)\frac{(b_{22}^+)^{m-1}}{(2m+1)!}\Bigl\{b_{22}^+
\Bigl[2(h^1-2)d^+_{12}d_{12}
 \nonumber\\
 && \quad +h^1- h^2-(d^+_{12})^2d_{12}^2\Bigr]
-2b_{12}^+d^+_{12}(h^1-d^+_{12}d_{12}-2)- b_{11}^+ (d^+_{12})^2
\Bigr\}b_2^{2m+1}
\nonumber\\
 &&  - \frac{m_2}{2} \sum_{m=0}
\sum_{l=0}\left(\frac{-2r}{m_2^2}
\right)^{m+l+1}\hspace{-1.0em}(4^m-1)\frac{(b_{22}^+)^{m+l-1}}{(2m)!}\Biggl\{\frac{m_1}{m_2}\frac{b_{22}^+b_1^+d^+_{12}}{(2l+2)!}b_2
 \nonumber\\
 && \hspace{1.0em}
+b_{12}^+d^+_{12}\left[\frac{(h^2 - h^1 + 2d^+_{12}d_{12}+
b_2^+b_2 +2}{(2l+1)!}  - \frac{b_2^+b_2}{(2l+2)!}\right]
 \nonumber \\
 && \hspace{1.0em}
-\frac{\Bigl(b_{11}^+(d^+_{12})^2 +b_{22}^+\bigl[h^2-h^1+b_2^+b_2
\bigr]\Bigr)}{(2l+1)!} \Biggr\}b_2^{2(m+l)+1}
\nonumber\\
 &&  + \frac{m_2}{2} \hspace{-0.2em}\sum_{m=0}
\sum_{l=0}\left(\frac{-2r}{m_2^2}
\right)^{m+l+1}(4^m-1)\frac{(b_{22}^+)^{m}}{(2m)!}
\frac{(b_{22}^+)^{l}}{(2l+1)!}\times\nonumber\\
 &&  \qquad \times d_{12}^+\left\{h^2-h^1+b_2^+b_2 +
d^+_{12}d_{12} +2\right\}d_{12}b_2^{2(m+l)+1}
\nonumber\\
 &&  - \frac{m_2}{2} \sum_{m=0}
\sum_{l=0}\left(\frac{-2r}{m_2^2}
\right)^{m+l+1}(4^m-1)\frac{(b_{22}^+)^{m}}{(2m)!}\frac{(b_{22}^+)^{l}}{
(2l+1)!}  \ \widehat{t}'_{12}\ d^+_{12} b_2^{2(m+l)+1},
 \label{l2oscnn2}\\
 \widehat{l}^{\prime}_{12} &= &
\frac{1}{4}\sum_{m=1}\left(\frac{-2r}{m_2^2}
\right)^{m}(b_{22}^+)^{m-1}d^+_{12}\left\{\frac{(h^1+h^2-2)}{(2m)!}+
\frac{b_2^+b_2}{(2m+1)!}\right\} b_2^{2m},
 \label{l12oscnn2}
\\
\widehat{l}^{\prime }_{22}  & = & - 2
\widehat{l}^{\prime}_{12}d_{12} -
\sum_{m=0}\left(\frac{-8r}{m_2^2} \right)^{m}
\frac{(b_{22}^+)^m}{m_2^2(2m+2)!}\bigl\{m_0^2-r
h^2(h^2-3)\bigr\}b_2^{2(m+1)}\nonumber\\
\hspace{-1.0em} && -\frac{1}{2}\sum_{m=1} \left(\frac{-8r}{m_2^2}
\right)^{m}
(b_{22}^+)^{m-1}b_2^+\left\{\frac{1}{2}\frac{(2h^2-1)}{(2m+1)!} +
\frac{b_2^+b_2}{(2m+2)!}\right\}b_2^{2m+1} \nonumber
\\
 &&  + \frac{1}{2}\sum_{m=0}
\left(\frac{-2r}{m_2^2}
\right)^{m+1}\hspace{-1.0em}(4^{m}-1)(b_{22}^+)^{m-1}
\left\{{\frac{b_{12}^+d^+_{12}}{(2m+2)!}}\bigl[2(h^1-
2)-2d^+_{12}d_{12}\bigr]
b_2\right.  \nonumber\\
 &&   \  + {\frac{b_{22}^+}{(2m+2)!}} \Bigl[h^2-h^1
-2d^+_{12}d_{12}(h^1-2)+ (d^+_{12})^2d_{12}^2\Bigr]b_2 +
{\frac{b_{11}^+(d^+_{12})^2 b_2}{(2m+2)!}}\nonumber\\
 && \left. \
 -{\frac{b_2^+d^+_{12}d_{12}}{(2m+1)!}} \right\}b_2^{2m+1} \ + \ \frac{m_1}{2m_2}\sum_{m=0}
\left(\frac{-2r}{m_2^2}
\right)^{m+1}{\frac{(b_{22}^+)^{m-1}b_1^+d^+_{12}}{(2m+1)!}}(4^{m}-1)
b_2^{2m+1}\nonumber\\
 &&
 +
\frac{1}{2}\sum_{m=0}\sum_{l=0}\left(\frac{-2r}{m_2^2}\right)^{m+l+1}
\hspace{-1.0em}(4^m-1)\frac{(b_{22}^+)^{m+l-1}}{(2m+1)!}\Biggl\{
\frac{m_1}{m_2}\frac{b_{22}^+b_1^+d^+_{12}}{(2l+2)!} b_2-
\frac{b_{11}^+(d^+_{12})^2}{(2l+1)!}
 \nonumber\\
 && \
+\frac{b_{22}^+}{(2l+1)!}\bigl(d^+_{12}d_{12}+1\bigr)
\bigl(h^1-h^2-b_2^+b_2-d^+_{12}d_{12}\bigr)\nonumber\\
 && \ + b_{12}^+d^+_{12}\left[\frac{(h^2 - h^1 +
2d^+_{12}d_{12} + b_2^+b_2+2)}{(2l+1)!} - \frac{b_2^+b_2
}{(2l+2)!} \right]
\Biggr\}b_2^{2(m+l+1)}\nonumber\\
 &&  +
\frac{1}{2}\sum_{m=0}\sum_{l=0}\left(\frac{-2r}{m_2^2}\right)^{
m+l+1}\hspace{-1.0em}(4^m-1)
\frac{(b_{22}^+)^{m+l}}{(2m+1)!(2l+1)!}\ \widehat{t}'_{12} \
d^+_{12}b_2^{2(m+l+1)}
 . \label{l22oscnn2}
\end{eqnarray}
Note, the additional parts $o^{\prime}_I(B,B^+)$ as the
 formal power series  in the oscillators $(B,B^+)$ do not obey the
 usual properties,
\begin{equation}
 \left(l^{\prime }_{lm}\right)^+\neq l^{\prime +}_{lm}\,,
 \qquad
\left(t^{\prime }_{12}\right)^+\neq t^{\prime +}_{12},\qquad
\left(l^{\prime }_{0}\right)^+\neq l^{\prime }_{0}\,, \qquad
\left(l^{\prime }_{m}\right)^+\neq l^{\prime +}_{m}\,,\ l\leq m,
\label{hermcong}
\end{equation}
if one should use the standard rules of Hermitian conjugation for
the new creation and annihilation operators, $(B_a)^+ = B_a$.
Restoration the proper Hermitian conjugation properties for
$o'_I$, is achieved by changing the scalar product in
$\mathcal{H}'$ as follows,
\begin{eqnarray}
\langle{\Phi}_1|\Phi_2\rangle_{\mathrm{new}} =
\langle{\Phi}_1|K'|\Phi_2\rangle\,, \label{newsprod}
\end{eqnarray}
for any vectors $|\Phi_1\rangle, |\Phi_2\rangle$ with some
non-degenerate operator $K'$. This operator is determined by the
condition that all the operators of the algebra  must have the
proper Hermitian properties with respect to the new scalar
product,
\begin{equation}\label{HCproperty}
 \langle{\Phi}_1|K'E^{- \prime\alpha}|\Phi_2\rangle =
\langle{\Phi}_2|K'E^{\prime\alpha}|\Phi_1\rangle^* ,  \
\langle{\Phi}_1|K'G^{\prime }|\Phi_2\rangle =
\langle{\Phi}_2|K'G^{\prime }|\Phi_1\rangle^*,
\end{equation}
for   $(E^{\prime\alpha};E^{-\prime\alpha}) = (l^{\prime }_{lm},
t^{\prime }_{12}, l^{\prime }_{m}; l^{\prime +}_{lm}, t^{\prime
+}_{12}, l^{\prime +}_{m})$, $G^{\prime } = (g^{\prime i}_0,
l^{\prime }_0)$. The relations (\ref{HCproperty}) lead to
definition the operator $K'$, Hermitian with respect to the
standard scalar product $\langle\, |\, \rangle$, in the form
\begin{eqnarray}
\label{explicit K} K'=Z^+Z, \qquad
Z=\sum_{(\vec{n}_{lm},\vec{n}_s)=(\vec{0},\vec{0})}^{\infty}
\left|\vec{N}(2)\rangle_V\right.\frac{1}{(\vec{n}_{lm})!(\vec{n}_{s})!}\langle
0|\prod_{r=1}^2b_r^{n_r}{d}_{12}^{p_{12}}\prod_{l,m \geq
l}{b}_{lm}^{n_{lm}},
\end{eqnarray}
where $(\vec{n}_{lm})! ={n}_{11}!{n}_{12}!{n}_{22}!$,
$(\vec{n}_{s})! = {n}_{1}!{n}_{2}!{p}_{12}!$ and the normalization
${}_V\langle0|0\rangle_V = 1$ is supposed.

The theorem 2 has the same consequences as ones from theorem 1
which concern, first, the flat limit of the algebra
$\mathcal{A}'(Y(2), AdS_d)$ and therefore a new representation for
Lie algebra $\left(T^{\prime 2} \oplus T^{\prime 2*}\oplus
l'_0\right)(0) + \hspace{-1em} \supset sp(4)$. Second, modulo the
oscillator pairs $b_{m2}, b^+_{m2}, b_2, b^+_{2}, d_{12},
d^+_{12}, m=1,2$,  the obtained representation coincides with one
for totally symmetric HS fields quadratic algebra
$\mathcal{A}'(Y(1), AdS_d)$ in \cite{BurdikNavratilPasnev}.

The set of the equations which compose the results of the Theorems
1, 2 represents  the general solution of mentioned in the
Introduction the second problem on the LF construction for
mixed-symmetry HS tensors on AdS spaces with given mass and spin
$\mathbf{s}=(s_1,s_2)$.

\section{Construction of Lagrangian Actions}\label{Lagrform}

To construct Lagrangian formulation  for the HS tensor
field of fixed generalized spin $\mathbf{s}=(s_1,s_2)$ we
should, initially, determine explicitly
 the composition law for the deformed algebra
$\mathcal{A}_{c}(Y(2), AdS_d)$ (for arbitrary $k\geq 2$ see \cite{BurdikReshetnyak}),
find BRST operator for the  non-linear algebra
$\mathcal{A}_{c}(Y(2), AdS_d)$ and, finally, reproduce  properly
gauge-invariant Lagrangian formulation for the basic bosonic field
$\Phi_{(\mu)_{s_1},(\nu)_{s_2}}$.

\subsection{Explicit form for the algebra $\mathcal{A}_c(Y(2),
AdS_d)$}\label{convalgg}

To this end as in the case of the algebra $\mathcal{A}'(Y(2),
AdS_d)$ of  $o'_I$ the only multiplication law for quadratic part
of the initial algebra $\mathcal{A}(Y(k), AdS_d)$ for k=2 is changed,
while its linear part is given by the same $sp(2k)$ algebra as for
the maximal Lie subalgebra
 for $\mathcal{A}(Y(2), AdS_d)$ and
$\mathcal{A}'(Y(2), AdS_d)$ with the same form of the commutators
$[O_a, O_I]$ for $O_a \in sp(4)$. From the Eqs.~(\ref{auxalg2}),
(\ref{conv-alg2}) and  Table~\ref{table} the  non-linear
 part of the algebra
 $\mathcal{A}_c(Y(2), AdS_d)$  can be
restored in the form of the Table~\ref{tablec},
\begin{table}[t]{\caption{The non-linear part of the converted algebra
$\mathcal{A}_c(Y(2), AdS_d)$.} \label{tablec}
\begin{center}
\begin{tabular}{||c||c|c|c||}\hline\hline
$[\,\downarrow\,,\to]$&
 $L_0$ &
$L^{ i}$ & $L^{i{}+}$  \\
\hline\hline $L_0$
    & $0$
   &
    $- r\hat{\mathcal{K}}^{ i+}_1$ & $r{\hat{\mathcal{K}}}^{ i}_1$\\
\hline $L^{j}$
   &   $ r\hat{\mathcal{K}}^{ j+}_1$
   & ${\hat{W}}^{ji}_b$  & ${\hat{X}}^{ji}_b$ \\
\hline $L^{ j+}$ &
   $-r\hat{\mathcal{K}}^{ j}_1$  &
   $-{\hat{X}}^{ij}_b$
   & $- {\hat{W}}^{ ji+}_b$  \\
   \hline\hline
\end{tabular}\end{center}
}
\end{table}
Here the functions $\hat{\mathcal{K}}^{
 i}_1$, ${\hat{W}}^{ ji}_b$,
 $\hat{X}^{ ij}_b$ (therefore  its Hermitian conjugated
 quantities $\hat{\mathcal{K}}^{ i+}_1$, ${\hat{W}}^{ ji+}_b$)
 are given as follows:
\begin{eqnarray}
{} \hat{W}^{ij}_b \hspace{-0.7em} & = & \hspace{-0.7em} 2r\epsilon^{ij}\Bigl\{\sum_l(-1)^l\bigl[G_0^{
l}-g_0^{\prime l}\bigr] L^{12} -
l^{\prime {12}}\sum_l(-1)^lG_0^{ l}-
\bigl[(T^{ 12}-t^{\prime 12})L^{ 11} - l^{\prime 11}T^{ 12}\bigr] \label{LiLj}\\
{}&&\qquad     +
(T^{+}_{12}-t^{\prime +}_{12})L^{22}- l^{\prime
22}T^{+}_{12}  \Bigr\},
 \nonumber\\
\hat{\mathcal{K}}^{ j}_1\hspace{-0.7em} & = & \hspace{-0.7em}
4\sum\nolimits_{i=1}^2\Bigl\{\bigl(L^{ ji+} - l^{\prime
ji+}\bigl)L^{ i}-l^{\prime i}L^{ ji+}\Bigr\} +
2\bigl(L^{j+}-l^{\prime j+}\bigr)G_0^{j} -
2\textstyle\bigl(g_0^{\prime j}+\frac{1}{2}\bigr)L^{ j+}  \nonumber\\
{}&& -2\Bigl\{\bigl[(L^{ 1+}- l^{\prime
1+})T^{ +}_{12}- t^{\prime +}_{12}L^{1+}\bigr]\delta^{j2}\delta^{i1} +
\bigl[(L^{
2+}-l^{\prime 2+})T^{ 12}- t^{\prime 12}L^{
2+}\bigr]\delta^{j1}\delta^{i2}\Bigr\} \label{L0Li+} ,
\\
 {} {\hat{X}}^{ij}_b
  \hspace{-0.7em} & = & \hspace{-0.7em}\Bigl\{L_{0}+ r\Bigl[ \hat{K}^{ 0i}_0 - 2g^{\prime i}_0 G_0^i +4 l^{\prime
ii+}L^{ii} +4 l^{\prime ii}L^{ii+}    +
\Bigl(\hat{\mathcal{K}}^{12}_0 - t^{\prime +}_{12}T^{ 12}- t^{\prime
12}T^{ +}_{12}
  \label{LiLj+b}\\
{}&& \qquad+4 l^{\prime 12+}L^{12} +4 l^{\prime 12}L^{12+}\Bigr)
\Bigr]\Bigr\}\delta^{ij}
      \nonumber\\
     \hspace{-0.7em} &  & \hspace{-0.7em}-
r\Bigl\{ 4\sum\nolimits_{l} \Bigl[(L^{jl+}-l^{\prime jl+})L^{ li}
- l^{\prime li}L^{ jl+} \Bigr]+ \textstyle\Bigl(\sum_l(G_0^{l
 j}-g_0^{\prime
 l})
  -2\Bigr)T^{ 12} - t^{\prime 12}\sum_lG_0^{l} \Bigr\}\delta^{i2}\delta^{j1}\nonumber\\
\hspace{-0.7em} &  & \hspace{-0.7em} - r\Bigl\{ 4\sum\nolimits_{l} \Bigl[(L^{jl+}-l^{\prime jl+})
L^{li} -l^{\prime li}L^{jl+}\Bigr]
 + \Bigl(T^{ +}_{ij}-t^{\prime +}_{ij}\Bigr)\sum_lG_0^{l}-
\Bigl(\sum_lg_0^{\prime l}+2\Bigr)T^{ +}_{12}\Bigl\}\delta^{i1}\delta^{j2}. \nonumber
\end{eqnarray}
The quantities $ \hat{K}^{ 0i}_0$, $\hat{\mathcal{K}}^{
12}_0$ above are the same as ones in
(\ref{Casimirsb}) for $k=2$, but expressed in terms of $O_I$.

Following to our experience from study of the (super)algebra
$\mathcal{A}_c(Y(1), AdS_d)$ in \cite{BurdikNavratilPasnev, BKL, adsfermBKR} when
in order to find exact BRST operator,  we choose the Weyl
(symmetric) ordering for quadratic combinations of $O_I$ in the
r.h.s. of the Eqs. (\ref{LiLj})--(\ref{LiLj+b}) as follows,
$O_IO_J = \frac{1}{2}(O_IO_J+O_JO_I)+ \frac{1}{2}[O_I\,,O_J]$. As
the result, the Table~\ref{tablec} for such kind of ordering must
contain the quantities ${\hat{W}}^{ij}_{b{}W}$,
$\hat{\mathcal{K}}^{ j}_{1{}W}$, ${\hat{X}}^{ij}_{bW}$ (and
${\hat{W}}^{ij+}_{b{}W}$, $\hat{\mathcal{K}}^{ j+}_{1{}W}$) which
read as
\begin{eqnarray}
{} {\hat{W}}^{ij}_{b{}W}\hspace{-0.7em} & = & \hspace{-0.7em}r\epsilon^{ij}\left\{\sum_l(-1)^l\mathcal{G}_0^{ l}
L^{12}
+ \mathcal{L}^{12}\sum_l(-1)^lG_0^{ l}-
\mathcal{T}^{ 12}L^{ 11} - \mathcal{L}^{ 11} T^{ 12}   +\mathcal{T}^{12+}L^{ 22} + \mathcal{L}^{ 22}T^{12+}
 \right\}
 \label{LiLjW}\\
\hat{\mathcal{K}}^{ j}_{1{}W} \hspace{-0.7em} & = & \hspace{-0.7em}
\Bigl\{2\sum_{l}\bigl[\mathcal{L}^{
jl+} L^{ l}+ \mathcal{L}^{ l}L^{ jl+}\bigr]
+ \mathcal{L}^{j+}G_0^{j}+ \mathcal{G}_0^{j}L^{ j+} - \bigl\{\mathcal{L}^{ 1+}T^{ +}_{12} + \mathcal{T}^{
+}_{12}L^{1+}\bigr\}\delta^{j2}\nonumber\\
 &&\qquad -\bigl[\mathcal{L}^{ 2+}T^{
12}+\mathcal{T}^{ 12}L^{ 2+}\bigr]\delta^{j1}\Bigr\} \label{L0Li+W} ,
\\
 {} {\hat{X}}^{ij}_{bW}
 \hspace{-0.7em} & = & \hspace{-0.7em}
\Bigl\{L_{0}+ r\Bigl[\mathcal{G}_0^{i}{G}_0^{i}
-2\mathcal{L}^{+}_{ii}L^{ii}
-2\mathcal{L}^{ii}L^{+}_{ii}+ \frac{1}{2} \bigl({\mathcal{T}}^{  12}T^{
+}_{12}+{\mathcal{T}}^{ +}_{12}T^{ 12}
   -4 \mathcal{L}^{+}_{12}L^{12} -4
\mathcal{L}^{12}L^{+}_{12}\bigr)\Bigr]\Bigr\}\delta^{ij} \nonumber\\
\hspace{-0.7em} &  & \hspace{-0.7em}
- r\Bigl\{ 2\sum\nolimits_{l}^2 (\mathcal{L}^{1l+}L^{
l2}+\mathcal{L}^{ l2}L^{ 1l+} ) +
\frac{1}{2}\sum_l^2\mathcal{G}_0^{
 l}T^{ 12} + \frac{1}{2}\mathcal{T}^{12}\sum_l^2{G}_0^{
 l} \Bigr\}\delta^{i2}\delta^{j1}
 \nonumber\\
\hspace{-0.7em} &  & \hspace{-0.7em}  - r\Bigl\{
2\sum\nolimits_{l}^2 (\mathcal{L}^{l2+} L^{1l}
+\mathcal{L}^{1l}L^{l2+}) + \frac{1}{2}\mathcal{T}^{ +}_{12}\sum_l^2{G}_0^{
 l}+ \frac{1}{2}\sum_l^2\mathcal{G}_0^{
 l}T^{ +}_{12}\Bigr\}\delta^{i1}\delta^{j2}
, \label{LiLj+bW}
\end{eqnarray}
with the notation  $\mathcal{O}_I$ for the quantity $\mathcal{O}_I
= ({O}_I - 2o'_I)$. Note, the  ordering quantities
(\ref{LiLjW})--(\ref{LiLj+bW}) do not contain linear terms (except
for $L_0$ in r.h.s of the last relation) as compared to
(\ref{LiLj})--(\ref{LiLj+b}).

 Thus, we derive the algebra  of converted operators $O_I$
underlying HS field subject to an arbitrary unitary irreducible
AdS group representation in AdS space with spin $\mathbf{s} =
(s_1,s_2)$ so that the problem now to find BRST operator for
$\mathcal{A}_c(Y(2), AdS_d)$.
\subsection{BRST-operator  for converted algebra $\mathcal{A}_c(Y(2), AdS_d)$}
\label{BRSTk2}

 The  non-linear algebra  now has not the form of closed
algebra because of  the operatorial functions
$F^{(2){}K}_{IJ}({o}',{O})$ in the Eq. (\ref{conv-alg2}) and as it
was shown in \cite{0812.2329} it leads to appearance of higher
order structural functions due to the quadratic algebraic
relations (\ref{LiLjW})--(\ref{LiLj+bW}) and their Her\-mi\-ti\-an
conjugates corresponding to those quantities
$F^{(2){}K}_{IJ}({o}',{O})$ for $i,j=1,2$. In ref.\cite
{0812.2329} it was found  new structure functions $F_{IJK}^{RS}(O)$
of the 3rd order in terminology of Ref.\cite{Henneaux}, implied by a resolution of the
Jacobi identities $[[{O}_I, {O}_J],{O}_K] + cycl. perm.
(I,J,K)=0$, as follows (\ref{sumcoeff}),
\begin{eqnarray}\label{Jid}
     \Bigl\{ F^{M}_{IJ} F^{P}_{MK}  + [F^{(2){}P}_{IJ},
   {O}_K] + cycl.perm.(I,J,K)\Bigr\} =
   F_{IJK}^{RS}\Bigl({O}_R\delta^P_S       -
   \textstyle\frac{1}{2} F^{P}_{RS})\Bigr) ,
   \end{eqnarray}
   for  $F^{M}_{IJ}\equiv \bigl(f^{M}_{IJ}+
   F^{(2){}M}_{IJ}\bigr)$.
The structure functions $F_{IJK}^{RS}({o}',{O})$ are antisymmetric
with respect to a permutation of any two of  lower indices
$(I,J,K)$ and  upper ones $R,S$ and exist   because of
nontrivial Jacobi identities for the $k(2k-1)=6$ triples $(L_i, L_j,
L_0)$, $(L_i^+, L_j^+, L_0)$, $(L_i, L_j^+, L_0)$.

 The construction of a BFV-BRST operator ${Q}'$ for  $\mathcal{A}_{c}(Y(2),
AdS_d)$ are considered in \cite{0812.2329} and has the general
form
\begin{eqnarray}\label{genQ'}
    {Q}'  = \mathcal{C}^I\bigl[{{O}}_I  + \textstyle\frac{1}{2}
    \mathcal{C}^{J}(f^{P}_{JI}+
   F^{(2){}P}_{JI})\mathcal{P}_{P}
    +\frac{1}{12}
    \mathcal{C}^{J}\mathcal{C}^{K}
    F^{RP}_{KJI}\mathcal{P}_{R}\mathcal{P}_{P}\bigr].
\end{eqnarray}
for the $(\mathcal{C}\mathcal{P})$-ordering for the ghost
coordinates
  $\mathcal{C}^I$ = $\{\eta_0$,
$\eta^i_G$ $\eta_i^+$, $\eta_i$, $\eta_{ij}^+$, $\eta_{ij}$,
$\vartheta_{12}$, $\vartheta^+_{12}\}$,  and their conjugated
momenta $\mathcal{P}_I$ = $\{{\cal{}P}_0$, ${\cal{}P}^i_G$ ,
${\cal{}P}_i$, ${\cal{}P}_i^+$, ${\cal{}P}_{ij}$,
${\cal{}P}_{ij}^+$, $\lambda_{12}^+$,
$\lambda_{12}\}$\footnote{they obey to independent nonvanishing
anticommutation relations
$
 \{\vartheta_{12},\lambda^+_{12}\}= 1,   \{\eta_i,{\cal{}P}_j^+\}= \delta_{ij}\,, \
 \{\eta_{lm},{\cal{}P}_{ij}^+\}= \delta_{li}\delta_{jm}\,,
 \{\eta_0,{\cal{}P}_0\}= \imath,
\{\eta^i_{\mathcal{G}}, {\cal{}P}^j_{\mathcal{G}}\}
 = \imath\delta^{ij}$,  possess the  standard  ghost number
 distribution,
$gh(\mathcal{C}^I)$ = $ - gh(\mathcal{P}_I)$ = $1$, providing the
property  $gh({Q}')$ = $1$, and have the Hermitian conjugation
properties of zero-mode pairs, $
 \left( \eta_0, \eta^i_{{G}},  {\cal{}P}_0,
{\cal{}P}^i_{G} \right)^+  =  \left( \eta_0, \eta^i_{G},  -
{\cal{}P}_0, -{\cal{}P}^i_{G}\right)$}. Explicitly, ${Q'}$ given
as,
 \begin{eqnarray}\label{explQ'}
   {Q'}\hspace{-0.5em}
& = \hspace{-0.5em}& Q'_1 + Q'_2 +
r^2\left\{\eta_0\sum\nolimits_{i,j}\eta_i\eta_j\varepsilon^{ij}\Bigl[
\textstyle \frac{1}{2}\sum_m\Bigl(G^m_0\bigl[\lambda_{12}
\mathcal{P}^{+}_{22}
 - \lambda_{12}^+
\mathcal{P}^{+}_{11}+ {i}
\mathcal{P}^{+}_{12}\sum_l(-1)^l\mathcal{P}_G^l\bigr]
   \right. \nonumber\\
&& \textstyle -
  {i}({L}^{+}_{11}\lambda_{12}^+-{L}^{+}_{22}\lambda_{12})
  \mathcal{P}_G^m + 4L^{mm}\mathcal{P}^{+}_{m2}
\mathcal{P}^{+}_{1m}\Bigr)
 - {L}^{+}_{12}\mathcal{P}_G^1\mathcal{P}_G^2  +2
L^{12}\mathcal{P}^{+}_{22}\mathcal{P}^{+}_{11}\Bigr]
\nonumber\\
&& \textstyle + \eta_0\sum_{i,j}\eta^+_i\eta_j\Bigl[
\sum_{m}\Bigl((-1)^m\frac{i}{2}G^m_0 \sum_l\mathcal{P}_G^l +
2(L^{+}_{22}\mathcal{P}^{22}-L^{11}\mathcal{P}^{+}_{11})\Bigr)
\lambda_{12}\delta^{1j} \delta^{2i} \nonumber\\
&& +
\varepsilon^{\{1j}\delta^{2\}i}\Bigl(\imath\textstyle\sum_m\Bigl[
\frac{1}{2}T^{12}\lambda_{12}^+ - 2 L^{12}\mathcal{P}^{+}_{12}
(-1)^m\Bigr]\mathcal{P}_G^m  +2\Bigl[
L^{12}\lambda_{12}-T^{12}\mathcal{P}^{12}\Bigr]\mathcal{P}^{+}_{22}
\nonumber\\
&&  +
2\Bigl[T^+_{12}\mathcal{P}^{12}-L^{12}\lambda_{12}^+\Bigr]\mathcal{P}^{+}_{11}
\Bigr) -T^{12}\Bigl[\mathcal{P}_G^1\mathcal{P}_G^2\delta^{2i}
\delta^{1j}+2\mathcal{P}^{11}\mathcal{P}^{+}_{22}\delta^{1i}\delta^{2j}\Bigr]
\nonumber\\
&& \left.\hspace{-1em} - \textstyle
2\sum_{m}(-1)^m\hspace{-0.2em}\Bigl[(G^m_0\mathcal{P}^{11}+\imath{L}^{11}\mathcal{P}_G^m)
\delta^{1i}\delta^{2j}- (G^m_0\mathcal{P}^{22} +
\imath{L}^{22}\mathcal{P}_G^m) \delta^{2i}\delta^{1j} \Bigr]
\mathcal{P}^{+}_{12}
  \Bigr] + h. c.\right\}\hspace{-0.2em}, \end{eqnarray}
with the standard form for linear $Q_1'$ and quadratic $Q_2'$
terms in ghosts $\mathcal{C}^I$ (see \cite{0812.2329} for
details). The Hermiticity of the
 nilpotent operator $Q'$ in total Hilbert space
 $\mathcal{H}_{tot}$ =
 $\mathcal{H}\bigotimes \mathcal{H}'\bigotimes \mathcal{H}_{gh}$
 is defined by the rule,
 \begin{equation}\label{tK}
  Q^{\prime +}K = K Q',\verb" for "K = \hat{1} \otimes K' \otimes
  \hat{1}_{gh},
\end{equation}
 with the operator $K'$ given in (\ref{explicit K}).
\subsection{Lagrangian formulation}\label{prLagrform}

Properly the construction of Lagrangians for bosonic  HS fields in
AdS${}_d$  space, can be developed by partially following the
algorithm of \cite{symint-adsmassless}, \cite{BKL}, (see, as well
\cite{BuchbinderReshetnyak}), which is a particular case of our
construction, corresponding to $s_2 = 0$. As a first step, we
extract the dependence of the BRST operator $Q'$ (\ref{explQ'}) on
the ghosts $\eta^i_{G}, {\cal{}P}^i_{G}$,
\begin{eqnarray}
\label{Q'} {Q}' \hspace{-0.4em} &=& Q + \eta^i_{G}(\sigma^i+h^i)+
\mathcal{B}^i \mathcal{P}^i_{G}
\end{eqnarray}
with some  inessential in later operators $\mathcal{B}^i$ and
   the BRST operator $Q$ which corresponds only to
the converted first-class constraints $\{O_I\} \setminus
\{G^i_0\}$,
\begin{eqnarray}
\label{Q} {Q}  &=& \textstyle
 \frac{1}{2}\eta_0L_0+\sum_{i}\eta_i^+L^i
+\sum_{l\leq m}\eta_{lm}^+L^{lm} + \vartheta^+_{12}T^{12}
 + \frac{\imath}{2}\sum_l\eta_l^+\eta^l{\cal{}P}_0
 \nonumber
\\
&& - \vartheta_{12}^+\sum\nolimits_{n}(1+\delta_{1n})\eta^{1n+}
\mathcal{P}^{n2}+ \vartheta_{12}^+
\sum\nolimits_{n}(1+\delta_{n2})\eta^{n2} \mathcal{P}^{+1n}+
\textstyle\frac{1}{2}\sum_{n }\eta^+_{n2}\eta^{1n}\lambda^{12}
\nonumber\\
&&
 - \textstyle\frac{1}{2}\sum\limits_{l\leq
m}(1+\delta_{lm})\eta^m\eta_{lm}^+\mathcal{P}^l  -
\bigl[\vartheta_{12} \eta^{+2} +\vartheta^+_{12} \eta^{+1}
\bigr]\mathcal{P}^2 \nonumber\\
&& + r\left\{
 \eta_0 \sum\nolimits_{i}\eta^+_i\Bigl[
 2\mathcal{L}^{ ii}\mathcal{P}_i^+ + 2\mathcal{L}^{ i+}\mathcal{P}_{ii}
  + \mathcal{G}^{i}_0  \mathcal{P}_{ i }+2
  \bigl(\mathcal{\mathcal{L}}^{12 }
  \mathcal{P}^{\{1+} +
\mathcal{L}^{ \{1+}\mathcal{P}_{12 }\bigr)\delta^{2\}i }
 \right.\nonumber \\
&&\left. \quad -\delta^{1i}\bigl(\mathcal{L}^{2}\lambda^+_{12}
 + \mathcal{T}^+_{12}\mathcal{P}_{2}\bigr) - \delta^{2i}
\bigl(\mathcal{L}^{ 1}\lambda_{12} +
\mathcal{T}_{12}\mathcal{P}_{1}\bigr)\Bigr]\right.\nonumber
\\
&& \left. - \textstyle\frac{1}{2}\sum_{i,j}\eta^+_i \eta^+_{j}
\varepsilon^{ij}\Bigl[\sum_m(-1)^m\mathcal{G}_0^m \mathcal{P}_{12}
- \bigl( \mathcal{T}_{12} \mathcal{P}_{11} +
\mathcal{L}^{11}\lambda_{12}\bigr)  + \mathcal{T}^+_{12}
\mathcal{P}_{22} + \mathcal{L}^{22}\lambda^+_{12} \Bigr]\right. \nonumber \\
&  & \left. +\textstyle 2\eta^+_i \eta_{j}\Bigl[\sum_{m}
\mathcal{L}^{jm+}\mathcal{P}^{im} -\frac{1}{8}\bigl(
\mathcal{T}^+_{12} \lambda_{12} +\mathcal{T}_{12}\lambda^+_{12}
\bigr)\delta^{ij} + \frac{1}{4}\sum_{m}\mathcal{G}_0^m
\lambda_{12} \delta^{j1}\delta^{i2} \Bigr] \right\}\nonumber\\
&&
 +
r^2\hspace{-0.3em}\left\{\eta_0\sum\nolimits_{i,j}\hspace{-0.3em}\eta_i\eta_j\varepsilon^{ij}\Bigl[
\textstyle \frac{1}{2}\sum_m\Bigl(G^m_0\bigl[\lambda_{12}
\mathcal{P}^{+}_{22}
 - \lambda_{12}^+
\mathcal{P}^{+}_{11}\bigr]
    + 4L^{mm}\mathcal{P}^{+}_{m2}
\mathcal{P}^{+}_{1m}\Bigr)  +2
L^{12}\mathcal{P}^{+}_{22}\mathcal{P}^{+}_{11}\Bigr]
 \right. \nonumber\\
&& \textstyle + \eta_0\sum_{i,j}\eta^+_i\eta_j\Bigl[
2\sum_{m}\bigl(
L^{+}_{22}\mathcal{P}^{22}-L^{11}\mathcal{P}^{+}_{11})\bigr)
\lambda_{12}\delta^{1j} \delta^{2i}
-2T^{12}\mathcal{P}^{11}\mathcal{P}^{+}_{22}\delta^{1i}\delta^{2j}
 \nonumber\\
&& \hspace{1em}+ 2\varepsilon^{\{1j}\delta^{2\}i}\Bigl(\bigl(
L^{12}\lambda_{12}-T^{12}\mathcal{P}^{12}\bigr)\mathcal{P}^{+}_{22}
  + \bigl(T^+_{12}\mathcal{P}^{12}-L^{12}\lambda_{12}^+\bigr)
\mathcal{P}^{+}_{11} \Bigr)
\nonumber\\
&& \left.\hspace{1em} - \textstyle
2\sum_{m}(-1)^m\hspace{-0.2em}\Bigl(G^m_0\mathcal{P}^{11}\delta^{1i}\delta^{2j}-
G^m_0\mathcal{P}^{22} \delta^{2i}\delta^{1j} \Bigr)
\mathcal{P}^{+}_{12}
  \Bigr] \right\}+ h. c.\footnotemark\,.
\end{eqnarray}\footnotetext{here, in writing the coefficients  depending on $o'_I, O_I$ we have
used the convention from Eqs.(\ref{LiLjW})--(\ref{LiLj+bW}), for
 $\mathcal{O}_I$} The generalized spin operator $\vec{\sigma} =
(\sigma^1,\sigma^2)$, extended by the ghost Wick-pair variables,
\begin{eqnarray}
\label{sigmai}
  \sigma^i = G_0^i - h^i   - \eta_i \mathcal{P}^+_i +
   \eta_i^+ \mathcal{P}_i + \sum_{
m}(1+\delta_{im})(
\eta_{im}^+{\cal{}P}^{im}-\eta_{im}{\cal{}P}^+_{im}) +
[\vartheta^+_{12} \lambda^{12} -
\vartheta^{12}\lambda^+_{12}](-1)^i\,,
\end{eqnarray}
commutes with  $Q$, $[Q,\sigma^i] =  0$. We choose a
representation for  Hilbert space $\mathcal{H}_{tot}$ coordinated
with decomposition (\ref{Q'}) such that the operators $ (\eta_i,
\eta_{ij}, \vartheta_{12}, \mathcal{P}_0, \mathcal{P}_i,
\mathcal{P}_{ij}, \lambda_{12}, \mathcal{P}^{i}_G)$ annihilate
vacuum vector $|0\rangle$, and suppose that the field vectors
$|\chi \rangle$ as well as the gauge parameters $|\Lambda \rangle$
do not depend on ghosts $\eta^{i}_G$,
\begin{eqnarray}
|\chi \rangle &=& \sum_n \prod\nolimits_{l}^2 ( b_l^+
)^{n_{l}}\prod\nolimits_{i\le j}^2( b_{ij}^+ )^{n_{ij}}( d_{12}^+
)^{p_{12}}( \eta_0^+ )^{n_{f 0}}  \prod^2_{i, j, l\le m, n\le o}(
\eta_i^+ )^{n_{f i}} ( \mathcal{P}_j^+ )^{n_{p j}} ( \eta_{lm}^+
)^{n_{f lm}} ( \mathcal{P}_{no}^+ )^{n_{pno}} \nonumber
\\
&&{}\times ( \vartheta_{12}^+)^{n_{f 12}} ( \lambda_{12}^+
)^{n_{\lambda 12}} |\Phi(a^+_i)^{n_{f 0} (n)_{f i}(n)_{p j}(n)_{f
lm} (n)_{pno}(n)_{f 12}(n)_{\lambda
12}}_{(n)_{l}(n)_{ij}p_{12}}\rangle \,. \label{chi}
\end{eqnarray}
The brackets $(n)_{f i},(n)_{p j}, (n)_{p no}$ in the
Eq.(\ref{chi}) means, for instance, for $(n)_{p no}$ the set of
indices $(n_{p 11},n_{p 12},n_{p 22})$. The  sum above is taken
over $n_{l}$, $n_{ij}$, $p_{12}$ and  running from $0$ to
infinity, and over the rest $n$'s from $0$ to $1$. The Hilbert
space $\mathcal{H}_{tot}$ is decomposed into a direct sum of
Hilbert subspaces with definite ghost number: $\mathcal{H}_{tot} =
\bigoplus_{k=-6}^6 \mathcal{H}_k$. Denote by $|\chi^k\rangle \in
\mathcal{H}_{-k}$, the state (\ref{chi}) with the ghost number
$-k$, i.e. $gh(|\chi^k\rangle)=-k$. Thus, the physical state
having the ghost number zero is $|\chi^0\rangle$, the gauge
parameters $|\Lambda \rangle$ having the ghost number $-1$ is
$|\chi^1\rangle$ and so on. For vanishing of all auxiliary
creation operators $B^+$ and ghost variables $\eta_0, \eta^+_i,
\mathcal{P}^+_i,...$ the vector $|\chi^0\rangle$ must contain only
physical string-like vector $|\Phi\rangle = |\Phi(a^+_i)^{(0)_{f
o} (0)_{f i}(0)_{p j}(0)_{f lm} (0)_{p{}no}(0)_{f 12}(0)_{\lambda
12}}_{(0)_l (0)_{ij}0_{12}}\rangle$,
\begin{eqnarray}\label{decomptot}
|\chi^0\rangle&=&|\Phi\rangle+  |\Phi_A\rangle ,\quad
|\Phi_A\rangle\Big|{}_{[B^+=\eta_0= \eta^+_i=
\mathcal{P}^+_i=\eta_{lm}^+ = \mathcal{P}_{no}^+ =
\vartheta_{12}^+ = \lambda_{12}^+=0]} = 0
\end{eqnarray}
One can show, using the part of equations of motion and gauge
transformations, that the vector $|\Phi_A\rangle$ can be
completely removed (see Ref.\cite{BurdikReshetnyak}).

The equation for the physical state ${Q}'|\chi^0\rangle=0$   and
the tower of the reducible gauge transformations,
$\delta|\chi\rangle$ = $Q'|\chi^1\rangle$, $\delta|\chi^1\rangle =
Q'|\chi^2\rangle$, $\ldots$, $\delta|\chi^{(s-1)}\rangle =
Q'|\chi^{(s)}\rangle$, lead to  relations:
\begin{eqnarray}
\label{Qchi}  Q|\chi\rangle & = &0,\qquad
\qquad(\sigma^i+h^i)|\chi\rangle=0, \qquad\left(\varepsilon,
{gh}\right)(|\chi\rangle)=(0,0),
\\
 \delta|\chi\rangle & = &Q|\chi^1\rangle, \qquad
(\sigma^i+h^i)|\chi^1\rangle=0, \qquad\left(\varepsilon,
{gh}\right)(|\chi^1\rangle)=(1,-1), \label{QLambda}
\\
 \ldots && \qquad\ldots  \qquad \ldots \nonumber\\
\delta|\chi^{s-1}\rangle &=& Q|\chi^{s}\rangle, \qquad
(\sigma^i+h^i)|\chi^{s}\rangle=0, \qquad \left(\varepsilon,
{gh}\right)(|\chi^{s}\rangle)= (s\, mod\, 2 ,-s). \label{QLambdai}
\end{eqnarray}
Here $\varepsilon$ means for Grassmann parity  and $s=6$ is the
maximal stage of reducibility for the massive bosonic HS field,
because of subspaces $H_{k} = \emptyset$, for all integer $k\leq -
7$. The middle set of equations in (\ref{Qchi})--(\ref{QLambdai})
determines the possible values of the parameters $h^i$ and the
eigenvectors of the operators $\sigma^i$. Solving spectral
problem, we obtain a set of eigenvectors,
$|\chi^0\rangle_{(n)_2}$, $|\chi^1\rangle_{(n)_2}$, $\ldots$,
$|\chi^{s}\rangle_{(n)_2}$, $n_1 \geq n_2 \geq 0$, and a set of
eigenvalues,
\begin{eqnarray}
\label{hi} \sigma_i | \chi \rangle_{(n)_k} = \Bigl(\textstyle
n^i+\frac{d-1-4i}{2} \Bigr) | \chi \rangle_{(n)_k},\quad -h^i =
n_i+\frac{d-1-4i}{2} \;, \ i=1,2\,,\ n_1 \in \mathbf{Z}, n_2 \in
\mathbf{N}_0\,.
\end{eqnarray}
It is easy to see that in order to construct Lagrangian for the
field corresponding to a definite Young tableau (\ref{Young k2})
the numbers $n_i$ must be equal to the numbers of the boxes in the
$i$-th row of the corresponding Young tableau, i.e. $n_i=s_i$.
Thus, the state $|\chi\rangle_{(s)_2}$ contains the physical field
(\ref{PhysState}) and all its auxiliary fields. We  fix some
values of $n_i=s_i$. After substitution: $h^i \to h^i(s_i)$
operator $Q_{(s_1,s_2)} \equiv Q_{\vert h^i \to h^i(s_i)} $, is
nilpotent on each subspace $H_{tot{}(s_1,s_2)}$ whose vectors
satisfy to the Eqs.(\ref{Qchi}) for (\ref{hi}). Hence, the
Lagrangian equations of motion (one to one correspond to
Eqs.(\ref{Eq-0b})--(\ref{Eq-3b}) for $k=2$),~a~sequ\-ence of
reducible gauge transformations  have the form
\begin{eqnarray}
\hspace{-1em}&& Q_{(s_1,s_2)}|\chi^0\rangle_{(s_1,s_2)}=0,  \quad
\delta|\chi^{s} \rangle_{(s_1,s_2)}
=Q_{(s_1,s_2)}|\chi^{s+1}\rangle_{(s_1,s_2)}, \ s =
0,...,5.\label{LEoM}
\end{eqnarray}
Analogously to  totally symmetric bosonic HS fields
\cite{symint-adsmassless}, \cite{BKL} one can show that Lagrangian
action for fixed spin $(n)_2=(s)_2$ is defined up to an overall
factor as follows
\begin{eqnarray}
\hspace{-1em}&& {\cal S}_{(s_1,s_2)} = \int d \eta_0 \;
{}_{(s_1,s_2)}\langle \chi^0 |K_{(s_1,s_2)} Q_{(s_1,s_2)}| \chi^0
\rangle_{(s_1,s_2)},\texttt{ for } |\chi^0\rangle\equiv
|\chi\rangle. \label{S}
\end{eqnarray}
where the standard scalar product for the creation and
annihilation operators is assumed with measure $d^dx\sqrt{|{g}|}$
over AdS space. The vector $| \chi^0 \rangle_{(s)_2}$   and the
operator $K_{(s)_2}$ in (\ref{S}) are respectively the vector $|
\chi \rangle$ (\ref{chi}) subject to spin distribution relations
(\ref{hi}) for HS tensor field
$\Phi_{(\mu^1)_{s_1},(\mu^2)_{s_2}}(x)$  and operator $K$
(\ref{tK}) where the  substitution
$h_i\to-(n_i+\frac{d-1-4i}{2})$  is done.
 The corresponding LF for bosonic
field with spin $\mathbf{s}$ subject to $Y(s_1,s_2)$ is a
reducible gauge theory of maximally $L = 6$-th stage of
reducibility.

One can prove that the  equations of motion (\ref{LEoM})
 indeed reproduces only the basic
conditions (\ref{Eq-0b})--(\ref{Eq-3b}) for HS
 fields with
given spin $(s_1,s_2)$ and mass.  Therefore, the resulting equations
of motion because of the representation (\ref{decomptot}) have the
form,
\begin{equation}\label{fincompEoM}
    L_0|\Phi\rangle_{(s)_2} = (l_0 + m_0^2)|\Phi\rangle_{(s)_2}, \
    (l_i, l_{ij}, t_{12})|\Phi\rangle_{(s)_2} = (0,0,0), \
    i\leq j.
\end{equation}
The above relations permit one to determine the parameter $m_0$ in
a unique way in terms of~$h^i(s_i)$,
\begin{equation}\label{m02}
  \textstyle  m_0^2  =
  m^2+r\Bigl\{\beta(\beta+1)+\frac{d(d-6)}{4}+
    \Bigl(h^1-\frac{1}{2}+2\beta\Bigr)\Bigl((h^1-\frac{5}{2}\Bigr)+
    \Bigl(h^2-\frac{9}{2}\Bigr)\Bigr\},
\end{equation}
whereas the values of parameter $m_1, m_2$ remain by completely
arbitrary and may be used to reach  special properties of the
Lagrangian for given HS field.

The general action (\ref{S}) gives, in principle, a direct recept
to obtain the Lagrangian for any component field
$\Phi_{(\mu^1)_{s_1},(\mu^2)_{s_2}}(x)$ from general vector $|
\chi^0 \rangle_{(s)_2}$ since the only what we should do it is a
computation of vacuum expectation values of products of some
number of creation and annihilation operators.

\section{Conclusion}
In the paper  we have derived the quadratic non-linear HS
symmetry algebra  for description of arbitrary integer HS  fields
on AdS-spaces with any dimensions and subject to $k$ row Young
tableaux $Y(s_1,\ldots, s_k)$. It is shown the difference of the
obtained algebras $\mathcal{A}(Y(k),AdS_d)$ for $k=2$,
$\mathcal{A}'(Y(2),AdS_d)$, $\mathcal{A}_c(Y(2),AdS_d)$
corresponding respectively to initial set of operators, their
additional parts and converted set of operators within additive
conversion procedure, is due to their pure non-linear parts, which
are, in turn, connected to the AdS${}_d$-radius $(\sqrt{r})^{-1}$
presence through the set of isometry AdS-space operators.

To obtain the algebras we start from an embedding of bosonic HS
fields into vectors of an auxiliary Fock space, treat the fields
as coordinates of Fock-space vectors and reformulate the theory in
such terms. We realize the conditions that determine an
irreducible AdS-group representation  with a given mass and
generalized spin in terms of differential operator constraints
imposed on the Fock space vectors. These  constraints generate a
closed non-linear algebra of HS symmetry, which contains, with the
exception of $k$ basis generators of its Cartan subalgebra, a
system of first- and second-class constraints. Above algebra
coincides modulo isometry group generators with its Howe dual
$sp(2k)$ symplectic algebra. The  construction  of a correct
Lagrangian description requires a deformation of the initial
symmetry algebra, into algebra $\mathcal{A}_c(Y(2),AdS_d)$
introducing the algebra $\mathcal{A}'(Y(2),AdS_d)$.

We have generalized the method of construction of Verma module
\cite{Dixmier} from the case of Lie (super)algebras
\cite{Liealgebra}, \cite{BurLeites}, \cite{BuchbinderReshetnyak}
 and for quadratic algebra  $\mathcal{A}'(Y(1),AdS_d)$ for
 totally-symmetric HS fields \cite{BurdikNavratilPasnev}, \cite{BKL}
  on to case of non-linear algebra underlying mixed-symmetric
  HS bosonic fields on AdS-space with two-row Young tableaux.
  The Theorem 1   presents our basic result in this relation.
We show that as the byproduct of Verma module derivation the
Poincare--Birkhoff--Witt theorem is valid in case of the algebra
under consideration, therefore providing the lifting of the Verma
module for Lie algebra $\mathcal{A}(Y(2),{R}^{1,d-1})$ [being
isomorphic to $\left(T^2 \oplus T^{2*}\right) + \hspace{-1em}
\supset sp(4)$] to one for quadratic algebra in a deformation
parameter $r$. Of course, the same it is expected to be true for
general algebra $\mathcal{A}'(Y(k),AdS_d)$, for which we suppose
to obtain the explicit form of Verma module  in the recursive
procedure manner by means of  new \emph{primary and derived
block-operators}, like $\hat{t}'_{12}$,  $\hat{t}^{\prime+}_{12}$.

   We have obtained the representation for the $15$ generators of the
   algebra $\mathcal{A}'(Y(2),AdS_d)$ over
   Heisenberg-Weyl algebra $A_6$ as the formal power series in
   creation and annihilation operators, which in case of flat
   space limit ($r=0$) takes the polynomial form, coinciding with
   earlier known results, at least  for $m=0$ \cite{flatbos} and appearing
   new one  for massive case \cite{0707.2181} and for $k=2$ in
   \cite{BuchbinderReshetnyak}. The Theorem 2
  finalizes our second basic result on solution of this Fock space realization
  problem for $\mathcal{A}'(Y(2),AdS_d)$ through Verma module construction approach.

On a base of BFV-BRST operator $Q'$ which was found in Ref.
\cite{0812.2329} exactly up to third degree in ghost coordinates,
for the nonlinear algebra $\mathcal{A}_c(Y(2),AdS_d)$ of 15
converted constraints $O_I$ by analyzing the structure of Jacobi
identities for them we present a proper construction of
gauge-invariant Lagrangian formulations for the bosonic HS fields
of given spin $\mathbf{s}=(s_1,s_2)$ and mass on AdS${}_d$ space.
 The corresponding
Lagrangian formulation is at most $6$-th stage reducible Abelian
gauge theory and is given by the Eqs.(\ref{LEoM}),({\ref{S}}). The
last relations  may be considered as the final result in solution
of the general problem to construct Lagrangian formulation for
non-Lagrangian initial AdS-group irreducible representations
relations which describe the bosonic HS field with two rows in
Young tableaux. One should be noted the unconstrained Lagrangians
for the free mixed-symmetry  HS fields with  two rows in Young
tableaux on a AdS background have not been derived until now in
both ``metric-like" and ``frame-like" formulations. These results
to be seen as the first step to interacting theory, following in
part to the research \cite{BuchbinderTsulaia},
\cite{BuchbinderTsulaia1}.

Among the directions for application of the obtained results we
point out the  developing of the unconstrained formulation with
minimal number of auxiliary fields for the basic HS field with two
and more rows in the Young tableaux analogously to formulation
given in \cite{quartmixbose}, \cite{quartmixbosemas} for totally
symmetric fields, which as well may be derived from the universal
Lagrangian formulation suggested in the paper.

  From a mathematical point of view the construction of the Verma
  module for the algebra $\mathcal{A}'(Y(2),AdS_d)$
  open  the possibility to study both its structure and search
  singular, subsingular vectors in it, so that it, in principle, will
  then permit to construct new (non-scalar) infinite-dimensional
  representations for given algebra. Besides, the above results
  permit to   definitely understand the problems of
  (generalized) Verma module   construction for HS symmetry algebras and
  superalgebras underlying HS bosonic and respectively fermionic fields on
  AdS-spaces subject to multi-row Young tableaux.

\ack The authors are grateful to J. Buchbinder, V. Dobrev,  V.
Krykhtin, P. Lavrov, E. Skvortsov, M. Vasiliev, Yu. Zinoviev,  for
useful discussions and E.Latini for correspondence. A.R. thanks
D.Francia for some illuminating comments and A.Galajinsky for
attraction the attention to the papers \cite{quartmixbose},
\cite{quartmixbosemas}. The work of $\check{C}$.B. was supported
in part by the GACR-P201/10/1509 grant and by the research plan
MSM6840770039. A.R. The work of A.R. was partially supported by
the RFBR grants, project No. 11-02-08343, project No. 12-02-00121
and by LRSS grant Nr.224.2012.2.
\appendix
\section*{Appendix}

\section{Proof of the Proposition}\label{proof}
 In this appendix we check the validity of the Proposition in the subsection~\ref{auxtheorem}.
 The proof is based on the
explicit derivation of the multiplication laws (\ref{addalg}) and
(\ref{conv-alg}), (\ref{sumcoeff}) for the sets $\mathcal{A}'$ of
the operators $o'_I$ and $\mathcal{A}_c$ of $O_I$. Namely, from
the right-hand-side of the relations (\ref{involrel}) we have
(with account for commutativity of $o_I$ with $o'_J$) the
equations to determine the unknown structural functions
${F}_{IJ}^K({o}',{O})$,
\begin{eqnarray}\label{exprOoo'} [\,{O}_I,{O}_J]=
[\,{o}_I,{o}_J]+[\,o_I',o_J'] = \sum_{m=1}^nf_{ij}^{K_1\cdots
K_m}\prod_{l=1}^{m}{o}_{K_l} + [\,o_I',o_J'].
\end{eqnarray}
Expressing in (\ref{exprOoo'}) the initial elements ${o}_{K_1},
\ldots , {o}_{K_n}$ through enlarged $O_I$ and additional $o'_I$
operators with use of $o'{O}$-ordering we obtain the sequence of
relations for each power of $o_K$,
\begin{eqnarray}\label{ok1}
  f_{IJ}^{K_1}{o}_{K_1}\hspace{-0.7em} &\hspace{-0.7em}=\hspace{-0.7em}&\hspace{-0.7em} f_{IJ}^{K_1}{O}_{K_1} - f_{IJ}^{K_1}
{o}'_{K_1}\,, \\
  f_{IJ}^{K_1K_2}{o}_{K_1}{o}_{K_2}
   & = & f_{IJ}^{K_1K_2}{O}_{K_1}{O}_{K_2} -
  (f_{IJ}^{K_1K_2}+ f_{IJ}^{K_2K_1}){o}'_{
K_1}{O}_{K_2} + f_{IJ}^{K_2K_1}{o}'_{K_1}
{o}'_{K_2} \label{ok1k2}\\
\cdots   &\cdots & \cdots \quad \cdots \quad \cdots \quad \cdots \quad \cdots \quad \cdots \quad\cdots \quad, \nonumber\\
  f_{IJ}^{K_1\cdots
K_n}\prod_{l=1}^{n}{o}_{K_l} & =  & f_{IJ}^{K_1\cdots
K_{n}}\prod\nolimits_{m=1}^{n}{O}_{K_m}
+\sum\nolimits_{s=1}^{n-1}(-1)^{s} f_{ij}^{\widehat{K_s\cdots
K_1}\widehat{K_{s+1}\cdots K_{n}}} \nonumber
\\
&\times& \prod\nolimits_{p=1}^{s}{o}'_{K_{p}}
\prod\nolimits_{m=s+1}^{n}{O}_{K_{m}}- (-1)^{
  n }f_{IJ}^{(n)K_{n}\cdots
K_1}\prod\nolimits_{s=1}^{n}{o}'_{K_s},\label{Ok1kn}
\end{eqnarray}
where the hats in the notation $f_{ij}^{\widehat{K_s\cdots
K_1}\widehat{K_{s+1}\cdots K_{n}}}$  for the quantities
$f_{ij}^{{K_s\cdots K_1}{K_{s+1}\cdots K_{n}}}$  means the set of
$\frac{n!}{s!(n-s)!}$ terms obtained through
$f_{ij}^{\widehat{K_s\cdots K_1}\widehat{K_{s+1}\cdots K_{n}}}$ by
the  symmetrization as it is explicitly shown in the Eqs.
(\ref{sumcoeff}). Above system (\ref{ok1})--(\ref{Ok1kn}) permits
one  to immediately establish, first, from the rightmost terms
above in (\ref{exprOoo'})--(\ref{Ok1kn}) that the set of ${o}'_I$
form the polynomial algebra $\mathcal{A}'$ of order $n$ subject to
the algebraic relations (\ref{addalg}). Second, the rest terms in
(\ref{exprOoo'})--(\ref{Ok1kn}) completely determine  the
structural functions $F^{(m){}K}_{IJ}({o}',{O})$, $m=1,\ldots, n$
in the form (\ref{expfunc}) and show that the set of $O_I$ indeed
determine  the non-linear algebra $\mathcal{A}_c$\footnote{The
algebraic relations (\ref{conv-alg}) for algebra $\mathcal{A}_c$
is differed from ones for polynomial algebra because of the
non-homogeneous character of the structural functions
$F^{(m){}K}_{IJ}({o}',{O})$ in $O_I$ due to presence of elements
$o'_I$}.

\end{document}